\documentclass{aa}  

\usepackage{graphicx}
\usepackage{txfonts}
 
\usepackage{graphicx}
\usepackage{txfonts}
\usepackage{natbib}
\usepackage{amsmath}
\usepackage{amsfonts} 
\bibpunct{(}{)}{;}{a}{}{,} 
\newcommand{\kms}{\,km\,s$^{-1}$}

\newcommand{\msun}{$\rm M_\odot$}

\usepackage[usenames]{color}

\def\niip{[\ion{N}{ii}]$\lambda6548$}
\def\niig{[\ion{N}{ii}]$\lambda6583$}

\def\siip{[\ion{S}{ii}]$\lambda6716$}
\def\siig{[\ion{S}{ii}]$\lambda6731$}

\def\oi{[\ion{O}{i}]$\lambda\lambda6300,6364$}
\def\oisx{[\ion{O}{i}]$\lambda6300$}
\def\oidx{[\ion{O}{i}]$\lambda6364$}

\def\ni{[\ion{N}{i}]$\lambda\lambda5197,5200$}
\def\nisx{[\ion{N}{i}]$\lambda5197$}
\def\nidx{[\ion{N}{i}]$\lambda5200$}

\def\oiiig{[\ion{O}{iii}]$\lambda5007$}
\def\oiiip{[\ion{O}{iii}]$\lambda4959$}

\def\niis{[\ion{N}{ii}]}
\def\oiiis{[\ion{O}{iii}]}

\def\Hbs{\ion{H}{$\beta$}}
\def\siis{[\ion{S}{ii}]}
\def\nis{[\ion{N}{i}]}
\def\ois{[\ion{O}{i}]}
\def\siis{[\ion{S}{ii}]}

\begin{document}

   \title{Optical emission lines in the most massive galaxies: morphology, kinematics and ionisation properties}


   \author{Ilaria Pagotto\inst{1} \and Davor Krajnovi\'{c}\inst{1} \and Mark den Brok\inst{1} 
   \and Eric Emsellem\inst{2} \and Jarle Brinchmann\inst{3,4} \and Peter M. Weilbacher\inst{1} 
   \and Wolfram Kollatschny\inst{5} \and Matthias Steinmetz\inst{1}
}

   \institute{Leibniz-Institute for Astrophysics Potsdam (AIP), An der Sternwarte 16, 14482 Potsdam, Germany \\ \email{ipagotto@aip.de} \and 
             ESO, European Southern Observatory, Karl-Schwarzschild Str 2, D-85748 
Garching bei Muenchen, Germany \and
Leiden Observatory, Leiden University, PO Box 9513, 2300 RA, Leiden, The Netherlands \and 
Instituto de Astrofísica e Ciências do Espaço, Universidade do Porto, CAUP, Rua das Estrelas, PT4150-762 Porto, Portugal \and
Institut f\"ur Astrophysik, Universit\"at G\"ottingen, Friedrich-Hund Platz 1, D-37077 G\"ottingen, Germany
}

   \date{Received xxx; accepted yyy}

  \abstract
{
In order to better characterize the upper end of the galaxy stellar mass 
range, the MUSE Most Massive Galaxies (M3G) Survey targeted the most massive 
galaxies (M $> 10^{12}$ \msun) found in the densest known clusters of galaxies 
at $z\sim0.046$. The sample is composed by 25 early-type galaxies: 14 BCGs, of 
which 3 are in the densest region of the Shapley Super Cluster (SSC), and 11 
massive satellites in the SSC. 
In the present work we aim at deriving the spatial distribution and kinematics
of the gas, and discussing its ionisation mechanism and origin in the optical 
wavelength range with MUSE data. 
We fit the continuum of the spectra using an extensive library of single stellar 
population models and model the emission lines employing up to three Gaussian 
functions.
In the M3G sample, ionized-gas was detected in five BCGs, of which one 
is in the densest region of the SSC, and six massive satellites in the 
SSC. Among these objects, \ois\ and \nis\ were detected in three BCGs 
and one satellite. The gas is centrally concentrated in almost all 
objects, except for two BCGs that show filaments and two massive satellites 
with extended emission. Moreover, the emission line profiles of three BCGs 
present red/blueshifted components. The presence of dust was revealed by 
analysing Balmer line ratios obtaining a mean $E(B-V)$ of 0.2-0.3. The 
emission-line diagnostic diagrams show predominately LINER line ratios 
with little contamination from star formation. In the M3G sample, the 
gas was detected in 80\% of fast rotators and 35\% of slow rotators. 
The orientations of stellar and gaseous rotations are aligned with 
respect to each other for 60\% of satellites and 25\% of BCGs. The 
presence of misalignments points to an external origin of the gas 
for three BCGs and two satellites. On the other hand, some of these systems 
are characterized by triaxial and prolate-like stellar rotation that 
could support an internal origin of the gas even in case of misalignments.}

   \titlerunning{Optical emission lines in the most massive nearby galaxies}
   
   \authorrunning{Pagotto et al.}
   
   \maketitle
%

\section{Introduction}


\begin{table*}
\caption{Details of the galaxies with gas.}
\centering
  	\label{tab:sample}
\begin{tabular}{ccccccccccc}
    \hline
    \noalign{\smallskip}
	\multicolumn{1}{c}{Galaxy} & 
	\multicolumn{1}{c}{Class} &
	\multicolumn{1}{c}{$r$} &
	\multicolumn{1}{c}{$M_{\rm gas}$} &
	\multicolumn{1}{c}{$E(B-V)$} &
    \multicolumn{1}{c}{n$^{\circ}$} &
    \multicolumn{1}{c}{$PA_{\rm g}$} & 
    \multicolumn{1}{c}{$PA_{\rm s}$} & 
    \multicolumn{1}{c}{$PA_{\rm p}$} & 
    \multicolumn{1}{c}{$\Delta PA_{\rm gs}$} & 
    \multicolumn{1}{c}{$\Delta PA_{\rm gp}$} \\ 
 & & & [10$^5$ \msun] & [mag] & & [$^{\circ}$] & [$^{\circ}$] & [$^{\circ}$]  & [$^{\circ}$] & [$^{\circ}$]\\
(1) & (2) & (3) & (4) & (5) & (6) & (7) & (8) & (9) & (10) & (11) \\
    \noalign{\smallskip} 
    \hline \hline
    \noalign{\smallskip}
PGC\,015524 & 1 & 50 & 29.0 & 0.1321 & 2 & $25.7\pm3.0$ & $116.5\pm12.6$ & $172.9\pm7.5$ & 90.8 & 147.2 \\
PGC\,046860 & 3 & 49 & 0.2* & 0.0566 & 1 & (...) & $56.0\pm1.8$  & $57.6\pm1.0$ & (...) & (...) \\
PGC\,047177 & 3 & 226 & 6.7 & 0.0542 & 2 & $107.4\pm2.3$   & $95.3\pm2.0$   & $98.4\pm1.3$ & 12.1 & 9.0 \\	
PGC\,047197 & 3 & 226 & 0.1* & 0.0501 & 1 & (...) & $24.2\pm29.7$  & $118.4\pm2.1$ & aligned & $\sim90$ \\	
PGC\,047202 & 2 & 226 & 4.5 & 0.0503 & 1 & (...) & $166.4\pm2.8$ & $172.2\pm2.5$ & (...) & (...) \\	
PGC\,047273 & 3 & 226 & 6.5 & 0.0502 & 1 & $28.7\pm2.3$   & $96.8\pm2.3$   & $93.6\pm2.2$ & 68.1 & 64.9 \\	
PGC\,047590 & 3 & 129 & 1.0 & 0.0549 & 1 & $102.9\pm3.8$  & $46.9\pm29.7$  & $132.4\pm3.8$ & 56.0 & 29.5 \\	
PGC\,049940 & 1 & 41 & 3.9 & 0.0663 & 3 & $54.4\pm0.8$   & $51.4\pm9.1$   & $51.7\pm1.9$ & 3.0 & 2.7  \\
PGC\,065588 & 1 & 66 & 1.3 & 0.0333 & 1 & $102.9\pm0.8$  & $136.1\pm2.3$  & $62.1\pm1.1$ & 33.2 & 40.8 \\	
PGC\,073000 & 1 & 66 & 44.8 & 0.0145 & 3 & $43.9\pm2.3$ & $66.6\pm5.0$   & $160.8\pm0.3$ & 22.7 & 116.9 \\
PGC\,097958 & 3 & 226 & 0.5* & 0.0649 & 1 & (...) & $45.4\pm1.5$  & $47.3\pm0.3$ & aligned & $\sim0$ \\

     \noalign{\smallskip}
    \hline
 \end{tabular}
\tablefoot{
Col(1): galaxy name. 
Col.(2): 1 = BCG, 2 = BCG in the SSC, 3 = ``satellite'' in the SSC. 
Col.(3): richness of the cluster from \citet{Laine2003}.
Col.(4): mass of the gas from H$\alpha$ flux. Values with * are upper limits.
Col.(5): foreground Galactic extinction from \citet{Schlegel1998}. 
Col.(6): number of Gaussian components employed to model the emission lines. 
Col.(7): gas kinematic position angle. 
Col.(8): stellar kinematic position angle from \citet{krajnovic2018}. 
Col.(9): $PA$ of the isophotal major axis from \citet{krajnovic2018}. 
Col.(10): absolute difference between column 6 and 7. Aligned = from 
a visual inspection, {\em PA$_{\rm g}$} is nearly aligned to {\em PA$_{\rm s}$}. 
Col.(11): absolute difference between column 6 and 8. For those galaxies with 
the ``aligned'' keyword in Col.(9), we report a rough estimation.}
\end{table*} 


The most massive early-type galaxies (ETGs) are rare objects in the local Universe 
and they represent the end point of the galaxy mass assembly. Most of them are giant 
ellipticals or lenticulars that stopped forming stars at $z\sim2$ \citep{Thomas2005, 
McDermid2015}, but at lower $z$ continued their evolution via dry mergers and minor 
accretion events of satellite systems (\citealt{vanDokkum2008}; \citealt{Thomas2014}), 
as also indicated by simulations (e.g., \citealt{Oser2012}; \citealt{Laporte2013};
\citealt{Wellons2015}, \citealt{Cooke2019}) and current analysis of stellar populations 
with integral-field spectroscopy (e.g. \citealt{Greene2015, Greene2019, Edwards2020}). 
According to present numerical simulations, these systems are likely the result of 
dissipation-less equal-mass mergers \citep{Lokas2014,Ebrova2015,Tsatsi2017,Li2018}. 
After $z\sim2$, massive ETGs grew from one major merger event around $z\sim1$, 
while BCGs are more likely to be built by more than one minor merger. 
In the local Universe, massive ETGs are characterized by low star formation 
rates \citep{Ford2013}, typically devoid of interstellar medium but 
with a large amount of X-ray emitting hot gas in their halo \citep{Ellis2006, 
Kormendy2009, Goulding2016}.
Most of them are located in dense environments, also as brightest members of 
groups or clusters (BCGs), while few of them are groupless \citep{Ma2014}. The 
galaxy population with stellar mass $M_{\star} \gtrsim 2 \times 10^{11}$ M$_{\odot}$ 
is dominated by slow rotators (\citealt{Cappellari2013}; \citealt{Veale2017}). 

The gas content of massive ETGs is poorly constrained given that for decades
these objects were thought to be characterized by a negligible amount of gas 
and the study of their evolution was mainly focused on stellar properties. 
Recently, the MASSIVE survey \citep{Ma2014} provided interesting insights on 
the gas distribution and properties in galaxies (isolated or in groups) more 
massive than 10$^{11.5}$ \msun\ within 108 Mpc. In particular, \citet{Davis2016} 
showed that 10\% of those objects have cold gas, which is at least a factor of 
two less then among less massive ETGs \citep{Young2011}. \citet{Goulding2016}
presented hot-gas properties of 33 early-type systems exploiting archival Chandra 
X-ray observations.
Warm ionised-gas 
was detected in 80\% of massive fast rotators but only in 28\% of massive slow 
rotators. In most of these massive galaxies this gas is centrally concentrated 
($\sim 61$\%), in some cases is extended and shows clear rotation ($\sim 29$\%) 
or patchy distribution ($\sim 3$\%). Its origin is driven by various physical 
processes as external accretion, cooling from the hot halo or AGN-driven outflow 
\citep{Pandya2017}.

In contrast, more studies focused on probing the gas content of BCGs, 
mostly due to their unique status within clusters. A small fraction of 
BCGs presents optical emission lines ($\sim 15$\%) and the presence of 
the gas is not correlated with the galaxy mass and cluster velocity 
dispersion \citep{Edwards2007}. They are characterized by various 
ionised-gas morphologies: centrally concentrated, extended or filamentary 
(\citealt{McDonald2012}). Recent gas accretion and star formation events 
are usually related to the X-ray cooling core status \citep{Bildfell2008, 
Edwards2020}. Indeed, most of BCGs at the center of the so called cooling 
flows (CF, \citealt{Fabian1994}; \citealt{Peterson2006}) clusters show 
optical emission lines (\citealt{Edwards2007, Cavagnolo2008, Tremblay2015}). 
This points to a close link between the central regions of BCGs and the galaxy 
cluster halo. In particular, the origin of H$\alpha$ filaments observed within 
BCGs is strongly related to soft X-ray emission \citep{McDonald2010}. The CF 
mechanism is balanced by AGN feedback that slows and eventually reheats the 
cooling gas in the cores of clusters, driving the evolution of these objects 
(\citealt{Hamer2016}; \citealt{Gaspari2018}) given that CFs can explain only 
some of the star formation activity in BCGs \citep{Groenewald2014}. 

Both massive ETGs and BCGs show low-ionization nuclear emission-line 
regions (LINERs) line ratios rather than ionisation from star-formation 
(\citealt{Linden2007}; \citealt{Loubser2013}; \citealt{Pandya2017}) that 
rarely characterizes these objects. Moreover, most of ETGs present reservoirs 
of diffuse ionized gas (DIG, \citealt{Papaderos2013}) over kpc scales with 
low-ionization emission-line regions (LIERs) ratios \citep{Belfiore2016,Gomes2016}. 
The DIG is likely to be mainly ionised by hot post-Asymptotic Giant Branch (AGB) 
stars \citep{Byler2019}.


In this work, we present the optical emission line analysis of the ETGs of 
the MUSE Most Massive Galaxies (M3G; PI: Emsellem; see \citealt{krajnovic2018}) 
observed with the Multi-Unit Spectroscopic Explorer (MUSE; \citealt{Bacon2010}). 
These galaxies are located in the densest known nearby clusters and suitable to 
better characterize the upper end of the galaxy stellar mass range 
($M_{\star} > 10^{12}$ M$_{\odot}$). This work provides key information on the 
gas properties of both massive satellites, for which the gas content is poorly 
constrained, and for BGCs of rich nearby clusters. 

The paper is organised as follows. 
In Section~\ref{sec:specanalysis} we present the M3G sample, previous results, 
and the data reduction of the MUSE spectra. In Section~\ref{sec:fitproc} we 
describe the spectral fitting procedure and in Section~\ref{sec:results} we 
present the spatially resolved gas distribution, kinematics, fluxes and 
extinction. We report the presence of misalignments between the orientation 
of the gaseous and stellar rotation in Section~\ref{sec:orientat} and we study 
the ionisation mechanism by exploiting the flux emission line ratios in 
Section~\ref{sec:BPT}. Finally, we discuss the different possible scenarios 
about the origin of the gas in each object in Section~\ref{sec:discussion} 
and we draw conclusions in Section~\ref{sec:conclusions}.


\begin{figure*}
\begin{center}
\includegraphics[scale= 0.58,angle=0]{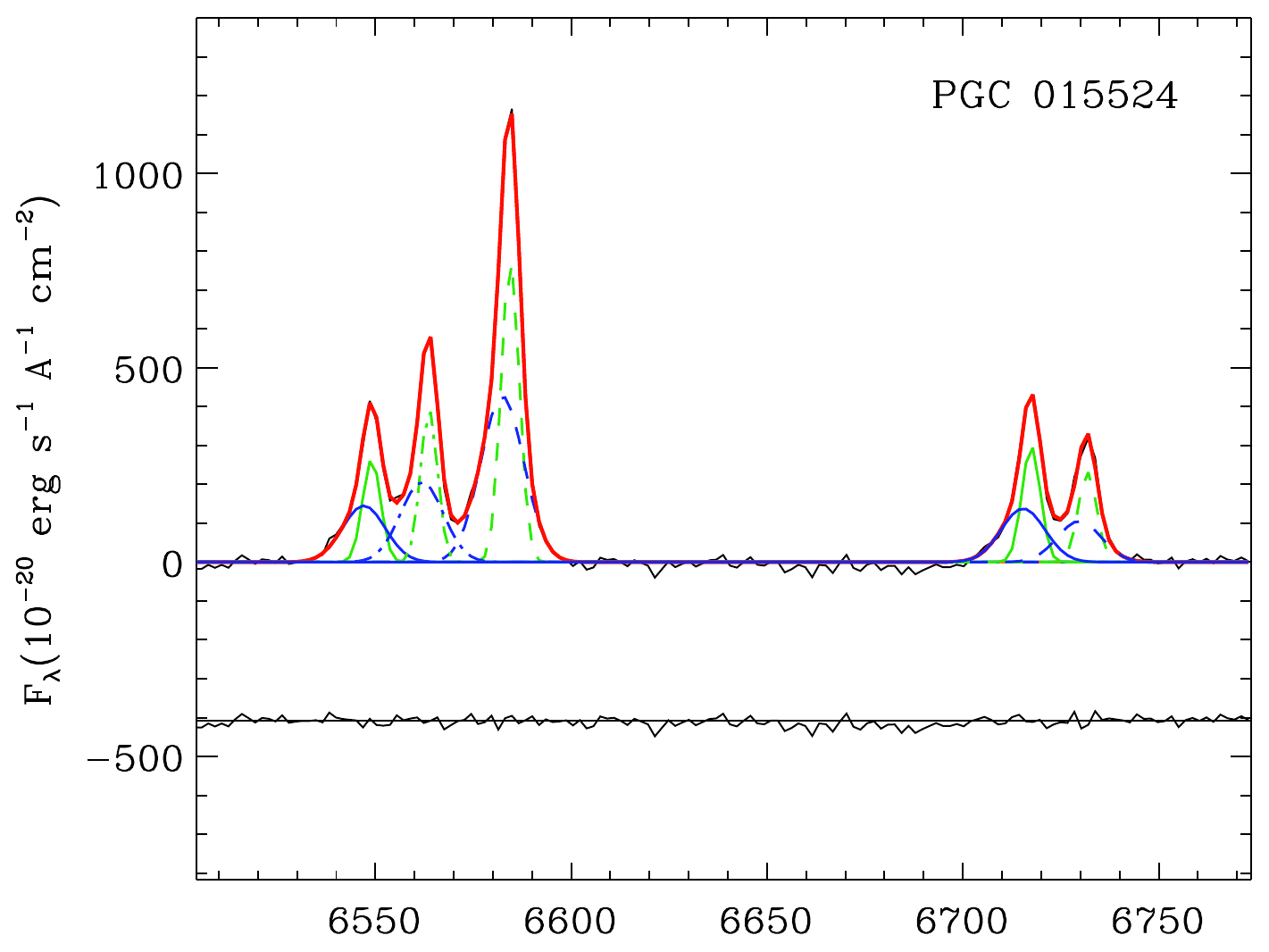} \quad
\includegraphics[scale= 0.58,angle=0]{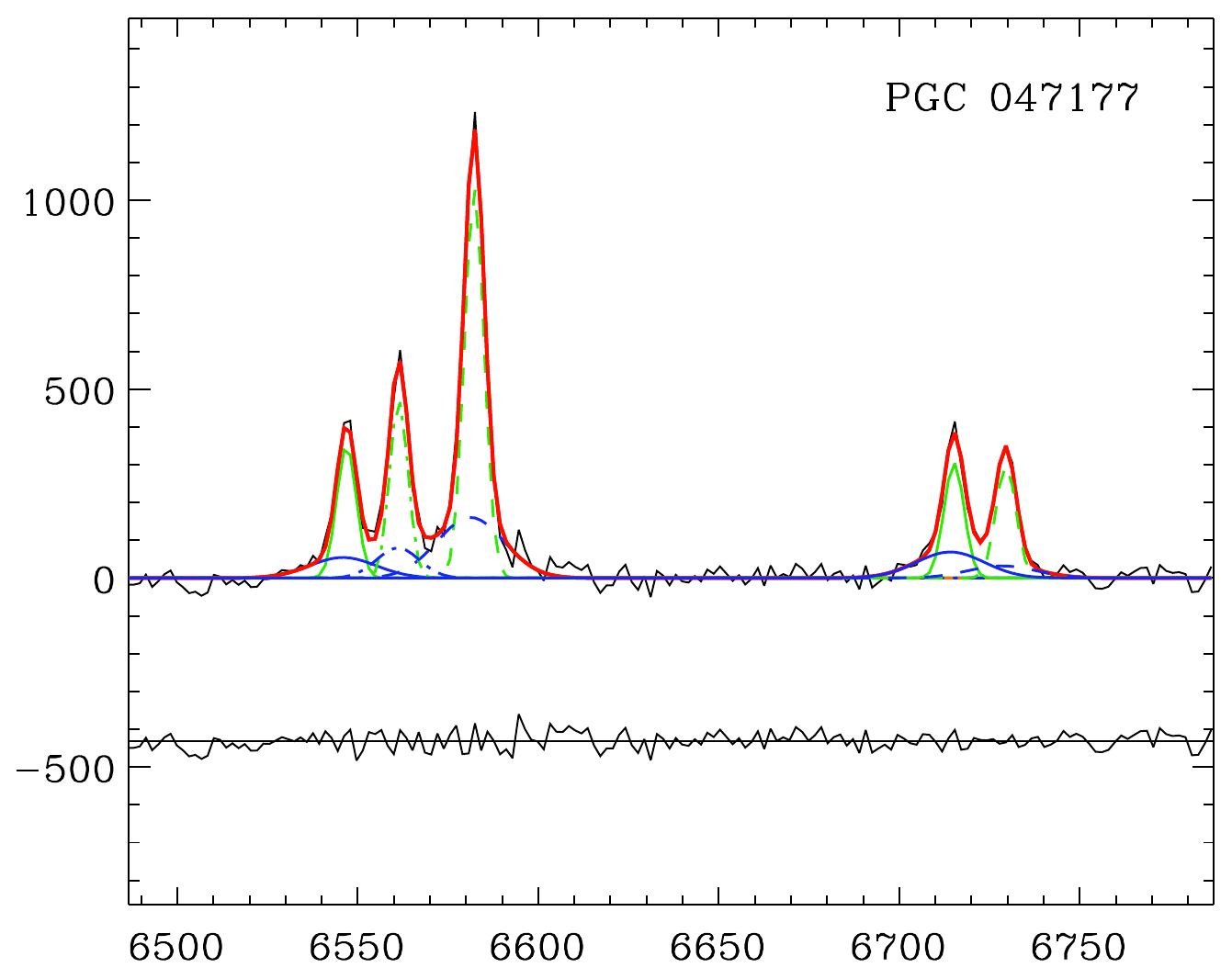} \\
\includegraphics[scale= 0.58,angle=0]{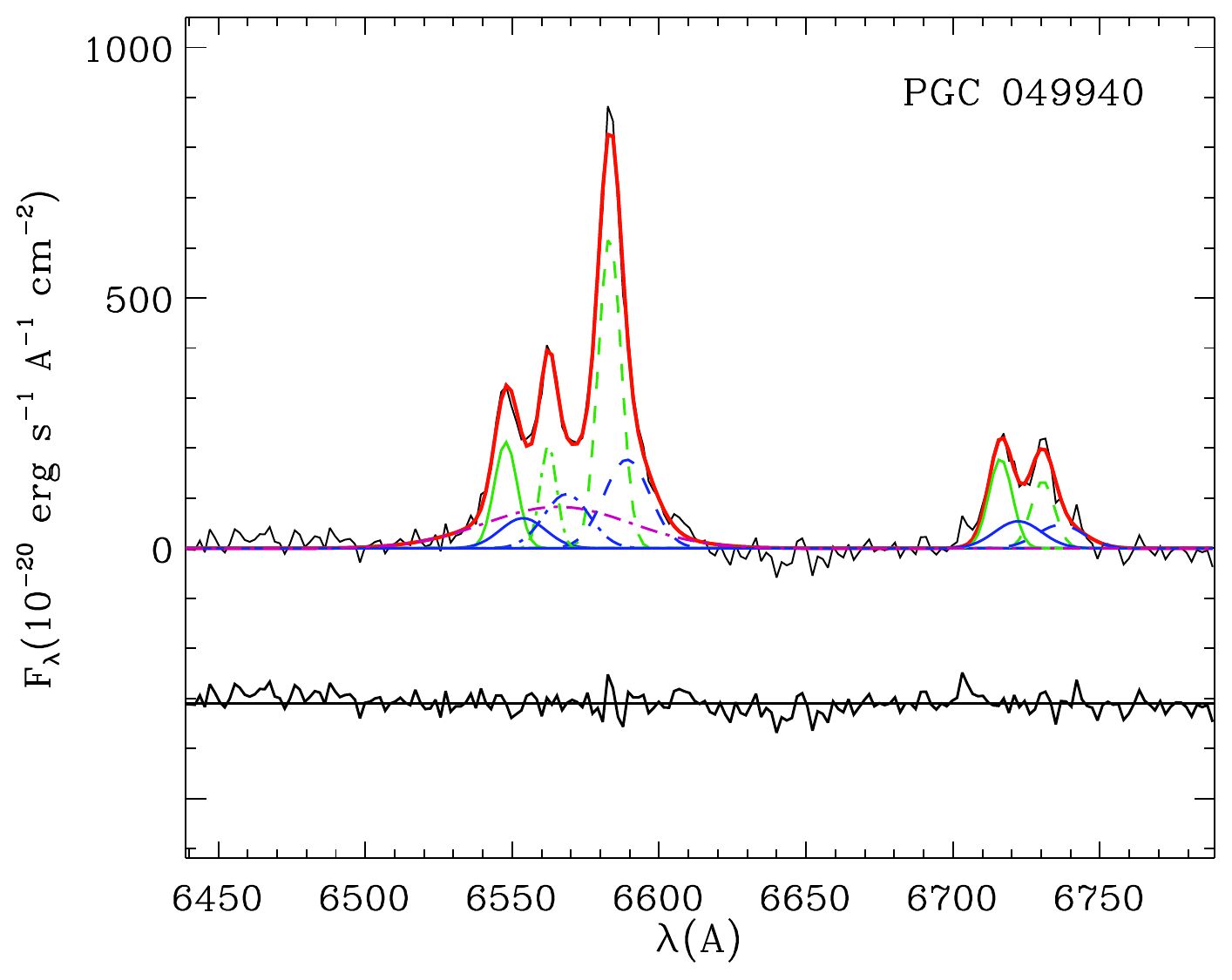} \quad
\includegraphics[scale= 0.58,angle=0]{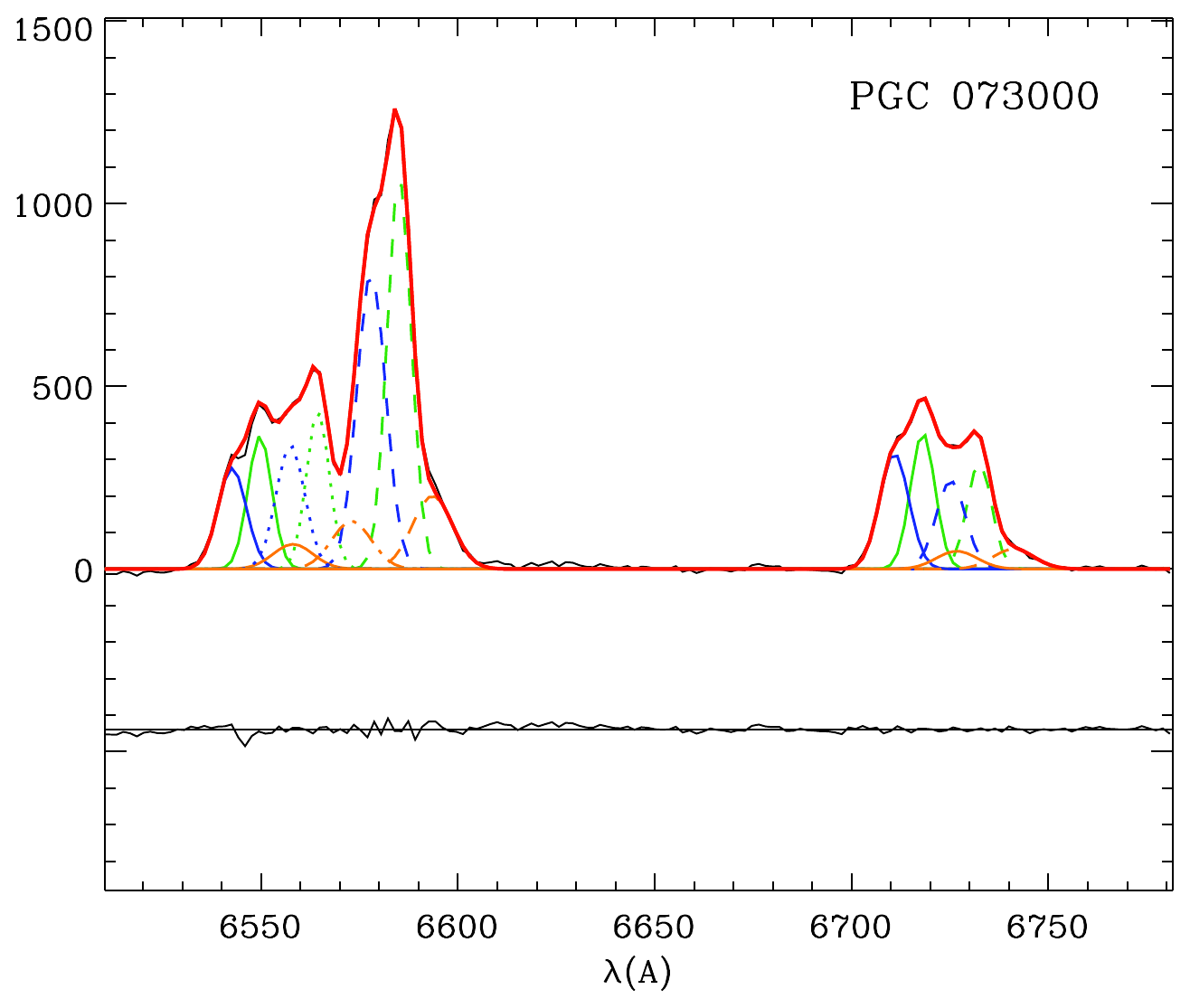} \\
\caption{MUSE continuum-subtracted spectra (black solid line) with best-fitting 
model (red solid line) obtained adopting first (green lines), blueshifted and second 
(blue lines), redshifted (orange lines) and, broad (purple line) components for \niis, 
H$\alpha$ and \siis\ lines. The Gaussian components of the same line are characterized 
by the same pattern. The residuals are defined as the difference 
between the observed and model spectrum. They are shifted to an arbitrary zero point for 
viewing convenience. The galaxy name is reported within each panel.}
\label{fig:lines}
\end{center}
\end{figure*}


\section{M3G sample and data reduction}
\label{sec:specanalysis} 

The M3G survey is composed of 25 massive ETG that were selected such that one MUSE 
pointing would cover approximately 2 $R_{\rm e}$ (based on 2MASS $R_{\rm e}$ values), 
with $M_{k} < - 25.7$ mag in the 2MASS $K_{s}$-band which ensures that their masses 
are $M_{\star} > 10^{12}$ M$_{\odot}$. They are characterized by a redshift range 
$0.037 < z < 0.054$, with a mean of $z = 0.046$. The galaxies are divided in two 
subsamples. The first one comprises 14 galaxies selected to be the BCGs in clusters 
richer than the Virgo Cluster. The other 11 are located in the core of the Shapley 
Super Cluster (SSC), centered around the three BCGs (from the BCG subsample) in the 
Abell \citep{Abell1989} clusters that make the core of the SSC (Abell 3562, Abell 
3558 and Abell 3556). About half of these galaxies are characterized by rotation 
around the major axis and stellar kinematic misalignment angles close to 90$^{\circ}$
\citep{krajnovic2018}. 

Unfortunately, the telluric sky lines fall in the wavelength range 
including H$\alpha$, \niis\ and/or \siis\ lines, depending on the 
redshift of each galaxy. As a consequence, an accurate (as much as 
possible) telluric correction is crucial to properly measure the  
weak emission lines. We produced again the reduced datacubes as in 
\citet{krajnovic2018} with the standard data reduction MUSE pipeline 
\citep{Weilbacher2020} except applying the sky telluric line correction 
step. The resulting datacubes were characterized by prominent telluric 
lines that were in turn removed with \textsc{molecfit} \citep{Smette2015}. 
This software computes a theoretical absorption model based on a radiative 
transfer code and an atmospheric molecular line database. The telluric 
spectra obtained with \textsc{molecfit} were then divided by each spectrum 
in the datacube. 

This analysis was aimed at obtaining a better telluric correction. 
Indeed, due to the lack of dedicated observations of telluric stars, 
\citet{krajnovic2018} conducted the telluric correction with the 
standard pipeline using standard stars. The residuals resulted from 
that correction were still strong and may significantly influence 
the measurements of weak emission lines. On the contrary, the residuals 
from \textsc{molecfit} procedure, while not completely absent, are weak 
and do not show a specific structure (i.e. mimicking emission-lines) allowing 
to clearly detect even the weakest emission lines in most of spectra. 

The cubes were Voronoi binned \citep{Cappellari2003}, as described in 
\citet{krajnovic2018}, by considering the signal-to-noise ratio (S/N) 
between 5500 and 5700 $\AA$. The target S/N was set to 50 for all galaxies.
The galaxy spectra were dereddened by correcting for Galactic extinction 
employing $E(B - V)$ values calculated from \citet{Schlegel1998} maps 
(see Table~\ref{tab:sample}), the extinction law from \citet{ODonnell1994} 
and $R_{V} = 3.1$. 


\section{Fitting procedure}
\label{sec:fitproc}

\subsection{Stellar continuum}

The stellar continuum was fitted with the Penalized Pixel Fitting (\textsc{pPXF},
\citealt{Cappellari2004}; \citealt{Cappellari2017}) method. We employed the MILES 
simple stellar population models \citep{Vazdekis2010} built with an Unimodal IMF
(with slope 1.3, equivalent to a Salpeter IMF) and Padova isochrones \citep{Girardi2000}, 
and convolved to the MUSE Line Spread Function (LSF) provided by \citet{Guerou2017}. 
The LOSVD was parametrized by Gauss-Hermite polynomials (\citealt{Gerhard1993};
\citealt{vanderMarel1993}) with the mean velocity $V$, velocity dispersion $\sigma$, 
and the higher order moments $h_3$ and $h_4$. Differently from \citet{krajnovic2018} 
who used a stellar optimal template (i.e., the linear combination of templates obtained 
from a pPXF fit on the spectrum resulted from the sum of all spectra within one effective 
radius) to fit each spectrum, we employed the entire library on each galaxy spectrum in 
order not to bias the flux emission line measurement. We exclusively included multiplicative
polynomials in the fitting process to account for flux calibration-related variations. 
Additive polynomials would risk biasing lines like H$\beta$ leading to the use of e.g., 
very young templates.


\subsection{Emission lines}
\label{sec:em}

We developed a procedure (see Sections.~\ref{sec:narrow} and \ref{sec:broad}) 
to properly fit both separated and blended lines with one or more velocity 
components. To model them we employed Gaussian functions convolved to the 
MUSE LSF \citep{Guerou2017}. For some galaxies the emission line profiles 
were simply fitted with a single Gaussian, while in other cases multiple 
Gaussians were required (see Sec.~\ref{sec:narrow}).


\subsubsection{Emission lines with one component}
\label{sec:narrow}

The stellar LOSVD obtained from \textsc{pPXF} fit was provided as input to Gas 
and Absorption Line Fitting (\textsc{GANDALF}, \citealt{Sarzi2017}) algorithm 
in order to model the emission lines with Gaussian functions. This procedure 
simultaneously fits stellar continuum and emission lines, allowing to properly 
detect even the weakest emission lines. At this stage, the reddening correction 
was not applied but multiplicative polynomials of the fourth order were used to 
adjust the continuum and correct bad flux calibration in the telluric spectral 
range not to overestimate the emission line flux. 

In order to identify which galaxies presented gas emission, we firstly fitted 
all the M3G Voronoi binned cubes with this technique. Gas emission lines were 
detected in 11 galaxies (see Table~\ref{tab:sample}). More details about the 
detection threshold can be found in Sec.~\ref{sec:extraction}. These objects 
show a variety of  emission line morphologies: (i) weak, (ii) strong and narrow 
or (iii) strong and broad emission lines. This procedure was successful 
for PGC\,047197, PGC\,047273, PGC097958 (satellites in Abell\,3558), PGC\,046860 
and PGC\,047590 (satellites in Abell\,3556 and Abell\,3562, respectively), 
PGC\,047202 and PGC\,065588 (BCGs of Abell\,3558 and Abell\,3716, respectively).
All of the emission lines of these galaxies were well modeled with 
one Gaussian function only. 


\begin{figure*}[!h]
\includegraphics[width=\textwidth, trim= 0cm 6.4cm 0cm 0cm, clip]{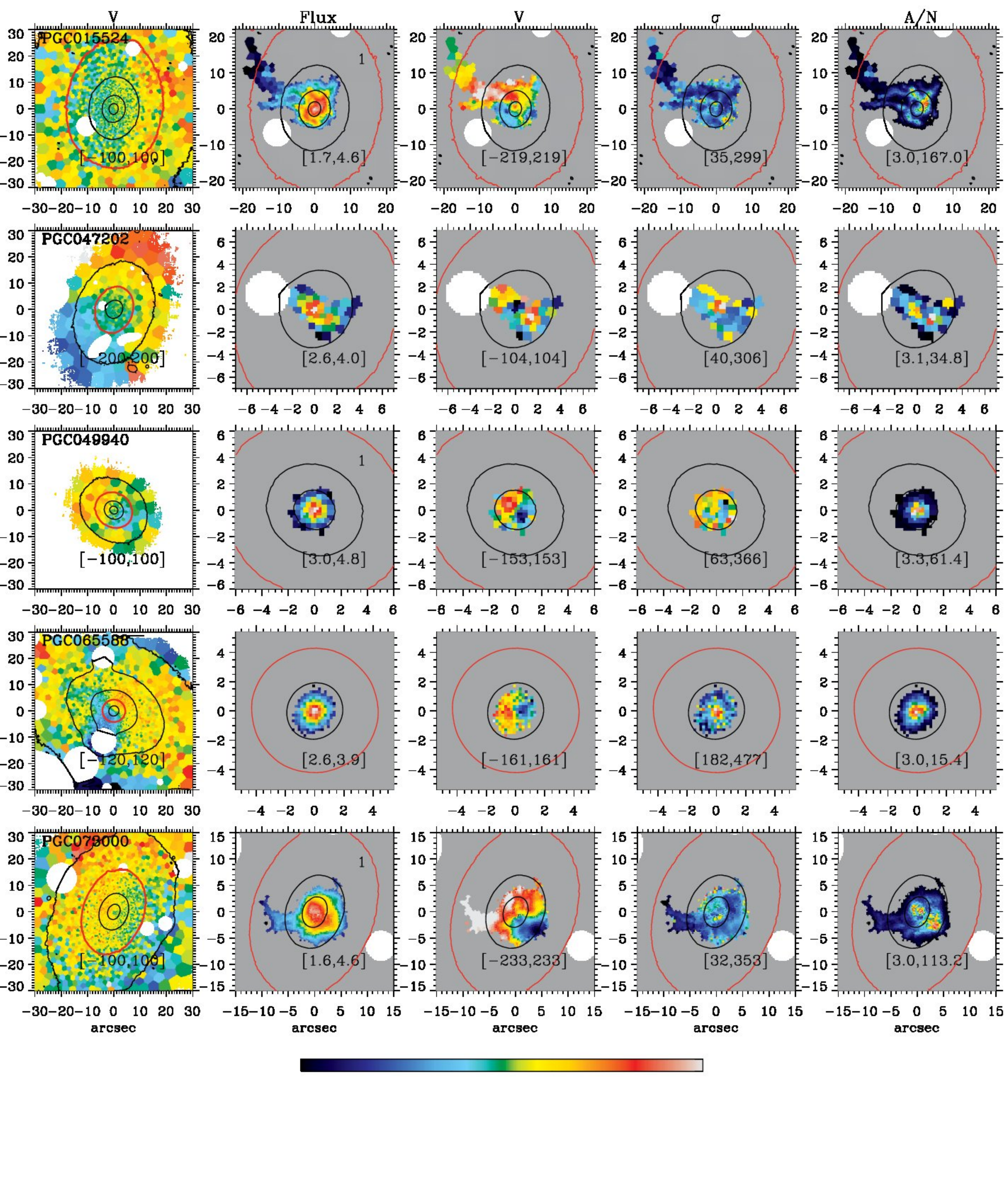} 
\caption{Spatially resolved maps of stellar velocities, \niis\ flux, 
kinematics and $A/N$ for the 5 BGCs with detected gas. Here the first 
component only is shown for those galaxies with emission lines with 
more than one component (those marked with 1 in the upper right corner 
of the flux maps of the gas). {\em Left to right}: stellar velocities 
in \kms, \niis\ flux in $\log$ 10$^{-20}$  erg~s$^{-1}$~cm$^{-2}$~arcsec$^{-2}$,
velocities in \kms, velocity dispersions in \kms\ and $A/N$ ratio. The maximum 
and minimum values adopted in the colorbar are reported in each panel between 
square brackets. Black and red contours are isophotes in steps of one magnitude. 
We enhanced the outermost isophote of the gas maps with red colour  in order to 
make the spatial region of the gas immediately visible on the stellar velocity 
panel. {\em Up to down}: Spatially resolved maps of PGC\,015524, PGC\,047202, 
PGC\,049940, PGC\,065588, and PGC\,073000.} 
\label{fig:BCGsmaps}
\end{figure*}

\begin{figure*}
\includegraphics[width=\textwidth, trim= 0cm 6.7cm 0cm 0cm, clip]{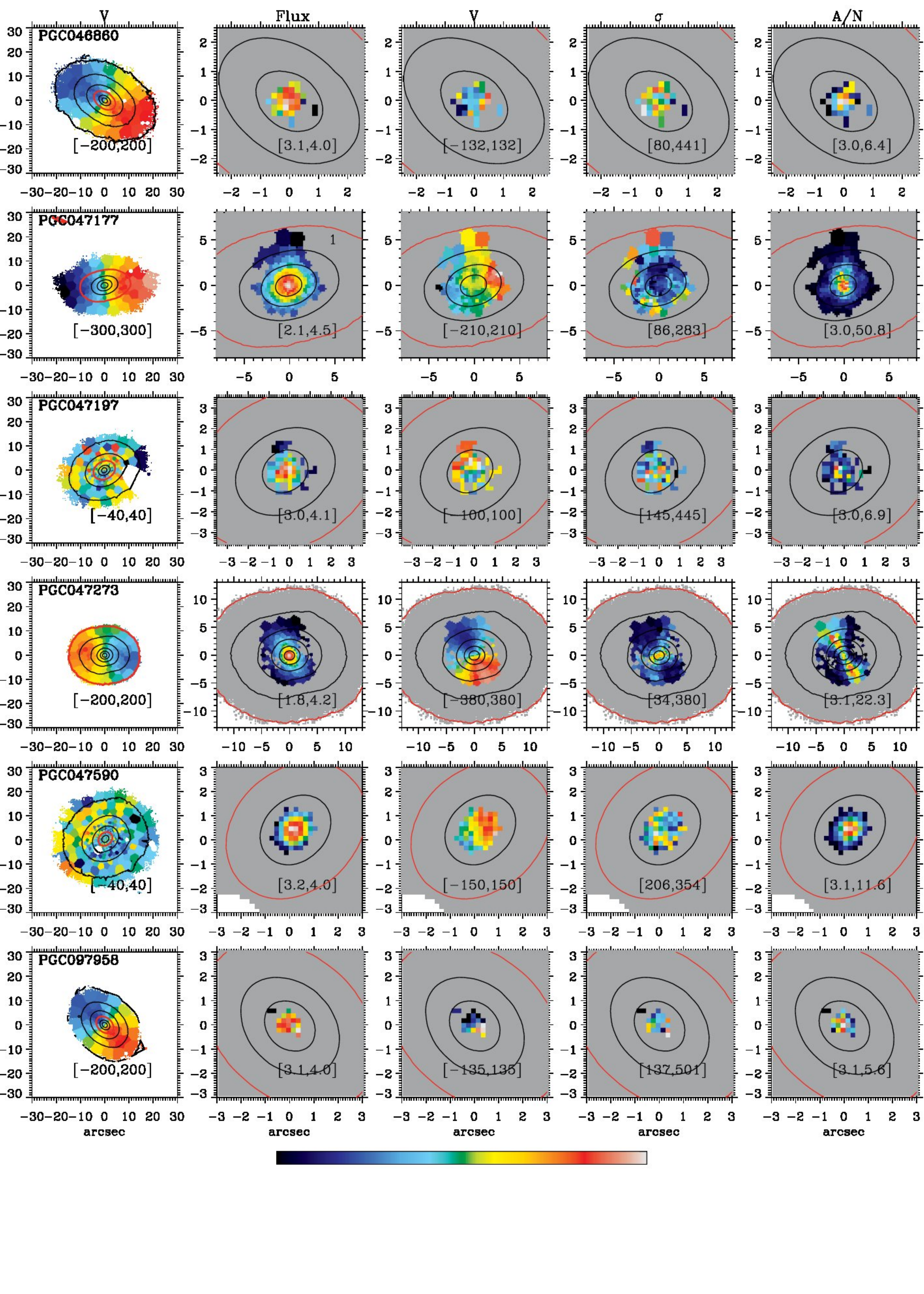} 
\caption{Same as in Fig.~\ref{fig:BCGsmaps} but for the 6 massive satellites. {\em Up to down}: 
Spatially resolved maps of PGC\,046860, PGC\,047177, PGC\,047197, PGC\,047273, PGC\,047590, 
and PGC\,097958.}
\label{fig:SATsmaps}
\end{figure*}


\subsubsection{Emission lines with more than one component}
\label{sec:broad}

The fitting procedure described in Section~\ref{sec:narrow} was not successful 
for most of spectra with blended emission lines characterized by more than one 
velocity component. This was the case of the emission lines in the central 
regions of PGC\,015524 (BCG of Abell\,0496), PGC\,049940 (BCG of Abell\,1836), 
PGC\,073000 (BCG of Abell\,4059) and PGC\,047177 (satellite in Abell\,3558).

Hereafter, we will refer to the gas velocity components that are rotating 
(regularly or affected by filaments) and in general narrower with respect 
to the others as ``first'' components. We will refer to the gas velocity 
components that are red/blueshifted with respect to the ``first'' one as 
``red/blueshifted'' components, and to the broad line region (BLR, $v \sim 
1000$ \kms) H$\alpha$ components as ``broad'' components. Only in the case 
of PGC\,047177, we will refer to the broader component as ``second'' component, 
since it is not red/blueshifted with respect to the first one. 

Since the emission lines with multiple components were all confined close to the 
galaxy center, we defined a circular spatial region that encloses all the spectra 
with multiple component line profiles. These spectra were fitted with a sum of 
Gaussian functions, after having subtracted the \textsc{pPXF} continuum. To carry 
out this analysis we developed a procedure based on \textsc{mpfit} 
IDL\footnote{Interactive Data Language is distributed by Harris Geospatial Solutions.} 
routine \citep{Markwardt2009}. The spectra outside the circular region were 
fitted with the procedure described in Section~\ref{sec:narrow} in order to properly 
detect the weakest emissions by simultaneously fitting stellar continuum and emission 
lines.

This new procedure allows 
$(i)$ to fit up to three Gaussian components for each line, 
$(ii)$ to tie the kinematics and flux ratios between lines, 
$(iii)$ to fix the line velocities of the nuclear bins (within 
the PSF) to the one of the central bin if required, 
$(iv)$ to measure the kinematics of a single line or doublet (e.g., 
H$\beta$ and \siis, respectively) and then keep it fixed during the 
fit of other lines to reduce the degeneration of parameters for blended 
lines (e.g., H$\alpha$ and forbidden lines, respectively),
$(v)$ to limit the range of variation of the values of all parameters in the fit, 
that is crucial to deal with possible exchange of Gaussian components during the 
fit of blended lines. 
We have tied the intensity ratios between the \oiiip\ and \oiiig , \nisx\ and \nidx , 
\oisx\ and \oidx , \niip\ and \niig lines (see \citealt{Osterbrock2006}), while 
keeping the ratio between the Balmer lines free to allow for the reddening correction. 
In order to reduce the degeneracy between parameters but still assure a good fit, the 
number of Gaussian components was kept as low as possible. More than one component 
was employed only in the case of complex non-Gaussian line profiles highlighted by 
large residuals (and high $\chi^2$/d.o.f. values) resulting from the one-component 
fit. The fit was visually inspected in case of strongly blended lines for which
a $\chi^2$-based selection of the number of Gaussians could lead to fits that are 
not physically justified. 

Fig.~\ref{fig:lines} shows some examples of the results obtained from the 
fitting procedure applied in the central circular region and in the following, we 
describe the adopted assumptions and number of Gaussian components employed for 
each galaxy.  

\paragraph*{PGC\,015524} Most of the ionised-gas emission lines within 
$\sim 5$\arcsec\ from the center were fitted with two Gaussian components 
(see Fig.~\ref{fig:lines}) while the \nis\ and \ois\ emission lines were 
properly fitted with one component only. The second component is blueshifted 
with respect to the first one of about 100 \kms. 
Due to degeneracy between parameters in blended lines, we forced each 
blueshifted component to have the same velocity and velocity dispersion for 
all the emission lines and in bins within the PSF we imposed the same 
velocities of the central bin. 
   
\paragraph*{PGC\,047177} The ionized-gas emission lines were fitted with 
two components (see Fig.~\ref{fig:lines}) while \nis\ with only one Gaussian 
component (no \ois\ was detected). In this case, the broader Gaussian component 
is not actually red/blueshifted (see Fig~\ref{fig:047177_2} in the Appendix A), so 
we will refer to it as ``second'' component. The velocity and velocity dispersion 
of Balmer and forbidden lines were fitted independently. To break the degeneracy 
between parameters, in the bins with two velocity components, firstly the \siis\ 
lines were fitted and then their kinematics were imposed to other forbidden lines 
during the fit. It was not possible to apply the same procedure to Balmer lines by 
previously fitting H$\beta$ because the H$\beta$ second component was not detected 
in almost all cases, probably due to extinction or some degeneracy between absorption 
and emission.  

\paragraph*{PGC\,049940} This galaxy shows very broad emission lines in the nuclear 
region, associated with the putative supermassive black hole ($3.61 \times 10^9$ 
\msun, \citealt{DallaBonta2009}). In addition, the presence of a redshifted
component, well visible in more external bins but detected also in the innermost 
ones thanks to the good $S/N$ ratio, makes this fit not so straightforward. 
The fit of the H$\alpha$ and \niis\ blended lines was quite challenging in the 
nuclear region where, in addition to first components, one redshifted component 
for each line plus a broad H$\alpha$ component (see Fig.~\ref{fig:lines}) were 
employed. The fit was less degenerate in the outer bins where the broad H$\alpha$ 
component disappears. In the nuclear bins, the \siis\ lines were fitted 
and their kinematics were fixed during the fit of the other forbidden lines. 
H$\beta$ was not exploitable for this aim since it was strongly extincted, 
probably due to the dust filament close to the nucleus, also visible in HST 
images.

\paragraph*{PGC\,073000} Most of the line profiles within 2\arcsec\ are 
characterized by three velocity components (see Fig.~\ref{fig:lines}). 
The first (and rotating) one is the more spatially 
extended, the second one is blue-shifted while the third one redshifted with 
respect to the first one. While the velocity dispersion of the blueshifted 
component was similar to the one of the first component ($\sim$ 150 \kms), 
the redshifted component was significantly broader ($\sim$ 250 \kms) and 
spatially confined to the central region. Due to the complexity of the fit and 
degeneracy between parameters in blended lines, we forced each velocity 
component to have the same velocity and velocity dispersion for all the 
emission lines. The three velocity components were strongly blended in the 
bins within a region comparable in size of the PSF ($\sim1$\arcsec) while 
more separated in the outer central region. H$\beta$ was not exploitable to 
reduce the degeneracy of parameters, since the redshifted component  
was not detected for most of bins. 


\section{Results}
\label{sec:results}

\subsection{Extraction}
\label{sec:extraction}

The values of the parameters of each emission line were extracted in 
order to build spatially resolved maps of fluxes, velocity, velocity 
dispersion and amplitude-to-noise ratio ($A/N$). Fig.~\ref{fig:BCGsmaps} 
and \ref{fig:SATsmaps} show the stellar velocity maps along with the 
maps pertaining to \niis\ emission, while the maps of all the detected 
emission lines are shown in the Appendix~\ref{sec:appendixA}.  

The $A/N$ provides us with an estimate of the accuracy with which one can 
measure the line parameters. Indeed, it strongly depends on how much the 
line sticks out with respect to the noise in the continuum or telluric 
residuals. As already mentioned before, the telluric residuals fall in 
the spectral range of \niis , H$\alpha$ and/or \siis\ lines, depending 
on the redshift. After the fit, the $A/N$ of these lines was re-calculated 
by considering the noise in the spectral range of the lines. 
As a result, we derived lower $A/N$ values with respect to those obtained 
from the \textsc{GANDALF} fit for those bins in which the telluric residuals 
were at the same level of the emission line signal. The bins with $A/N$ lower 
than 3 were discarded and considered as telluric residuals rather than actual 
emission lines. The same $A/N$ threshold was considered also for the other 
emission lines. Finally, the errors associated to the parameters were computed with
MonteCarlo simulations fitting (for each bin) 200 spectra with additional random 
noise and considering the $\sigma$ of the Gaussian distribution of values of 
parameters obtained from the fit.


\subsection{Gas distribution and kinematics}
\label{sec:detect}

In this section we describe the spatially resolved maps shown in Fig.~\ref{fig:BCGsmaps} 
and \ref{fig:SATsmaps} and Appendix~\ref{sec:appendixA}. We subtracted the stellar systemic 
velocity from the velocity maps of the gas. 

As already pointed out in Section~\ref{sec:em}, PGC\,015524, PGC\,047177, 
PGC\,049940, PGC\,073000 present lines with more than one velocity component, 
while PGC\,046860, PGC\,047197, PGC\,047202, PGC\,047273, PGC\,047590, PGC\,065588 
and PGC\,097958 lines with one detected velocity component only. 

The peak of the emission flux corresponds nearly or exactly to the photometric 
center in all the galaxies. Only for PGC\,015524 and PGC\,073000, there is no
clear emission peak but a more extended high flux spatial region, probably due 
to the perturbed gas distribution.

Considering the M3G sample, $\sim$28\% of BCGs and $\sim64$\% of massive 
satellites of the SSC contain ionised-gas. \ois\ and \nis\ were detected 
only in the 4 galaxies with multi-component line profiles. Overall, the 
strongest and most spatially extended emission is the one coming from 
\niis\ lines (see Fig.~\ref{fig:BCGsmaps} and \ref{fig:SATsmaps}), that 
are indeed the only ones detected in PGC\,046860, PGC\,047197 and PGC\,097958.  

In case of multiple velocity components, most of the flux is emitted from the 
first one. The ionised-gas of the first component is centrally concentrated in 
three of the BCGs (PGC\,047202, PGC\,049940 and PGC\,065588) while extended with 
outward filaments in the remaining two (PGC\,015524 and PGC\,073000). In particular, 
the filaments of PGC\,015524 are stretched towards the North-East direction with 
respect to the center and their morphology is the same for all emission lines, even 
if with different spatial extents. \ois\ and \nis\ follow the ionised-gas distribution, 
even if they are less spatially extended. PGC\,073000 presents one thick filament 
in the East direction clearly visible in the \niis\ and \siis\ lines, while barely 
detectable in Balmer and \oiiis\ lines and not at all present in \ois\ and \nis\
spatial maps. Moreover, a ``swelling'' appears in the south-west direction in all
of the maps. 

The gas is centrally concentrated in most of satellites (PGC\,046860, PGC\,047177, 
PGC\,047197, PGC\,047590 and PGC\,097958). 
In PGC\,047273, the gas appears to be distributed in a strongly tilted 
polar disk settled close to the minor axis. In the outskirts the disk-like 
structure bends in opposite directions towards the major axis plane, especially 
in the \oiiis\ and \siis\ spatially resolved maps. \niis\ and H$\alpha$ emissions 
are not spatially extended outwards as others due to the strong telluric residuals 
still present after the correction which prevented us from considering the most 
external bins as reliable. The $A/N$ spatial distribution is peculiar with lower 
values in the center that gets higher along and within the disk and decrease again 
outwards. 

Concerning the velocity field of the BCGs, the gas of the first component is 
regularly rotating in PGC\,049940 and PGC\,065588 while not showing rotation in 
PGC\,047202. The velocity structure in PGC\,015524 does not appear as a simple 
rotation pattern, not an unexpected feature considering the filamentary appearance 
of its distribution. In PGC\,073000, a more structured velocity pattern appears, 
while still far from being regular. 
The velocity dispersion has a clear peak in the center of PGC\,065588 (up to 
about 500 \kms), while it reaches values of about 300, 250 \kms for PGC\,047202 
and PGC\,049940, respectively, not showing a clear nuclear raise. In the case 
of PGC\,015524, the velocity dispersion presents a peak in the nucleus ($\sim 
200$ \kms) and is lower within the filaments (less than 100 \kms) with respect 
to their edges ($\sim 150$ \kms), suggesting that the gas is moving along the 
filaments generating more turbulent motions at their edges. The gas velocity 
dispersion map of PGC\,073000 shows elevated values near the center ($\sim 250$ 
\kms) and another rise in the south-west region ($\sim 150$ \kms). It is low 
in the filament (less than 100 \kms) suggesting that this object is in a later 
and more relaxed stage of gas accretion with respect to that in PGC\,015524. 


\begin{figure*}
\includegraphics[scale=0.2]{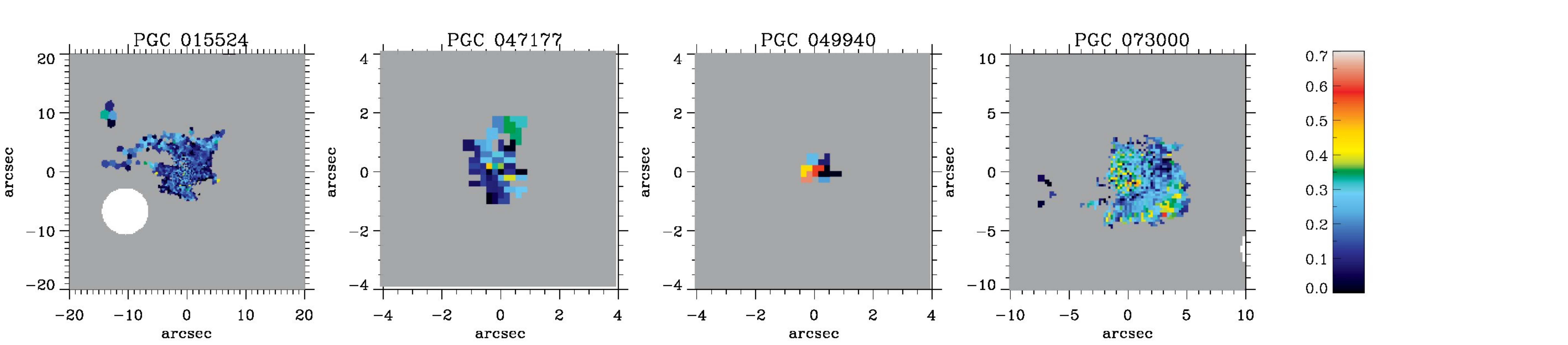} 
\caption{{\em Left to right}: spatially resolved extinction maps of PGC\,015524, 
PGC\,047177, PGC\,049940 and PGC073000 derived from the first line components. 
Their spatial extent is constrained by the detection of the $H\beta$ line.
The median errors fo the extinction are 0.10, 0.18, 0.15, and 0.10, respectively.} 
\label{fig:ext}
\end{figure*}


In satellites, the gas is clearly rotating in PGC\,047590, it seems to be 
in rotation in PGC\,097958 and PGC\,047197, while it does not show any rotation 
in PGC\,046860 . These 4 galaxies present a central peak in gas velocity 
dispersion ($\sim 300$ \kms). The gas rotation in PGC\,047177 is slightly 
perturbed, especially in the inner 2\arcsec . Looking at forbidden lines, 
the velocity dispersion is lower in the central region and increases going 
outwards. Finally, the polar gas disk in PGC\,047273 rotates nearly around 
the major axis and the velocity slightly decreases  in the outer 
regions where the disk is probably bending towards the major axis plane. 
The velocity dispersion is higher in the center ($\sim 300$ \kms) and 
decreases outwards (less than 100 \kms) as expected in a disk structure. 

Overall, the emission from the red/blueshifted components is more centrally 
concentrated with respect to the one from the first component. 
The blueshifted component of PGC\,015524 (Fig.~\ref{fig:015524_2}) is less 
spatially extended than the first one suggesting that the ionization source 
is located close to the centre. It shows a shallow velocity gradient that is 
not coupled to the one of the first component in that spatial region and high 
velocity dispersion values especially in the center that suggests the presence 
of biconical outflows \citep{Veilleux2005}. 
The second component of PGC\,047177 (Fig.~\ref{fig:047177_2}) is more confined 
with respect to the first one, it is characterized by a large velocity dispersion 
($\sim 400$ \kms\ for \niis) and a slightly negative velocity going outwards. 
The redshifted (of $ \sim 300$ \kms with respected to the first one) component 
of PGC\,049940 (Fig.~\ref{fig:049940_2}) is characterized by a velocity dispersion 
rising in the center up to nearly 500 \kms\ and may be associated with an AGN-driven
in/outflow of gas. It is more centrally concentrated than the first one even if the 
spatial scales are comparable. The broad ($ \sigma \sim 1000$ \kms) component 
(Fig.~\ref{fig:049940_3}) results from the BLR of the host galaxy. 
The blueshifted component of PGC\,073000 (Fig.~\ref{fig:073000_2}) may be 
interpreted as a large scale (up to nearly 4 kpc) outflow of gas at about 200 
\kms. The velocity dispersion increases going outwards where the gas velocity is 
slowing down. On the contrary, the redshifted component (Fig.~\ref{fig:073000_3}) 
is less extended ($\sim 1$ Kpc), broader and is approaching us at about 400 \kms. 

\subsection{Ionised gas masses}
\label{sec:masses}

We estimated the masses of the gas by using the same approach 
as in \citep{Sarzi2006}. We assumed a case B recombination, 
a temperature of T = $10^4$ K, an electron density of 
$n_{\rm e}$ = $10^2$ cm$^{-3}$, and $V_{\rm 3K}$ distances ($D$) from NED
obtained with $H_{0} = 70$ km~s$^{-1}$~Mpc$^{-1}$ (except the distance 
to PGC~097958 which is from Hyperleda). Since our targets do not exhibit 
significant ongoing star formation, we assumed a constant ionised gas 
density and use \citet[page 665]{Kim1989} formula to estimate the masses

\begin{equation*}\begin{split}
M~(\mathrm{M_{\odot}}) = 2.8 \times 10^2 & \left(\frac{D}{10~\mathrm{Mpc}}\right)^2 \\ & \times 
~\left(\frac{F(\mathrm{H\alpha})}{10^{-14}~\mathrm{erg~s^{-1}~cm^{-2}}}\right) 
\left(\frac{10^{3}~\mathrm{cm}^{-3}}{n_{\mathrm{e}}}\right)
\end{split}\end{equation*}

The masses were derived by considering 
the total flux of H$\alpha$ for each galaxy and the $V_{\rm 3k}$ 
distance reported in NED (for PGC\,097958 from HyperLeda). For 
PGC\,046860, PGC\,047197 and PGC\,097958 we provide an upper limit 
for the mass by summing the residuals of each bin around H$\alpha$ 
($50 \AA$) after having subtracted the stellar continuum. The 
resulting values are reported in Tab.~\ref{tab:sample}.

Concerning actual estimates, the amount of gas spans a range 
of $10^5$ - $5\times10^6$ \msun. Galaxies with filaments 
(PGC\,015524 and PGC\,073000) or spatially extended gas (PGC\,047273 
and PGC\,047177) clearly contain more gas than ones with a more centrally 
concentrated gas distribution (PGC\,047202, PGC\,047590 and PGC\,049940). 
The upper limits for the mass of the gas are of the order of $10^4$ \msun. 
Except for PGC\,015524 and PGC\,073000, the amount of ionised-gas in
our galaxies is of the same order of the gas contained in MASSIVE 
galaxies \citep{Pandya2017}, even if not directly comparable since 
we used H$\alpha$ rather than H$\beta$ flux.

\subsection{Reddening}
\label{sec:reddening}

We inspected the dust content exploiting the flux ratio between first 
components of Balmer lines. We used the Calzetti extinction law 
\citep{Calzetti2000} to derive the flux attenuation as a function of 
the wavelength at any value of $E(B-V)$. We assumed for the Balmer 
decrement a case B recombination with a predicted H$\alpha$/H$\beta$ 
ratio of 2.86 \citep{Osterbrock1989}. 

The reddening correction was performed on PGC\,015524, PGC\,047177, PGC\,049940 
and PGC\,073000. The Balmer decrement was safely measured in those bins with 
$A/N$ of H$\beta$ larger than 3. It was not possible to inspect the dust 
content in the other sample galaxies since H$\beta$ was not detected at all. 
In Fig.~\ref{fig:ext} are shown the $E(B-V)$ spatially resolved maps.

As can be seen, in all of these objects some extinction is present. PGC\,015524
shows $E(B-V)$ values up to 0.3 that are lower on average with respect to ones 
found by \citep{Loubser2013}. Notably, the $E(B-V)$ is a 
bit higher in filaments. The extinction map of PGC\,047177 reveals the presence 
of some dust in the center with $E(B-V)$ lower in the south-east than in 
the north-west spatial bins. Interestingly, the region with higher extinction 
corresponds to the spatial extent of a dip in the velocity dispersion values 
visible both in H$\alpha$ and forbidden emission and discrepant velocities with 
respect to the background gas structure (see the velocity maps). This could point 
to the presence of a small central gas and dust disk which counter-rotates with 
respect to the stars and gas at larger scales, even if it is a marginal feature 
given the shallow difference ($\sim 50$ \kms) in the central velocities. 
PGC\,049940 is characterised by high values of $E(B-V)$ close to the nucleus. 
From HST/ACS images it is possible to see that there is a optically thick dust 
filament in the central part of the BCG. PGC\,073000 shows an average value of 
$E(B-V)$ of about 0.2 even if it raises up to 0.4 close to the center and in 
the south-east region, corresponding to a X-ray arm \citep{Choi2004}. 


\section{Stellar vs gas orientation}
\label{sec:orientat} 

We derived the global kinematic position angle of the gas ({\em PA$_{\rm g}$}) 
and compared it to the one of the stars ({\em PA$_{\rm s}$}) provided by 
\citet{krajnovic2018}. We measured {\em PA$_{\rm g}$} from the spatially 
resolved velocity map of the \niig\ emission line (of the first \niig\ 
component in the case of line profiles with multiple components) employing 
the procedure {\tt fit\_kinematic\_pa}\footnote{https://www-astro.physics.ox.ac.uk/~mxc/software/}
(see \citealt{Krajnovic2006}, appendix C) and obtained the results reported
in Tab.~\ref{tab:sample}. All the {\em PA$_{\rm g}$} were measured from North 
to East. 

This analysis was not successful for PGC\,046860, PGC\,047197 and PGC\,097958 
since the spatial extent of the gas is limited to a few central bins. From 
a visual inspection, we inferred that in PGC\,047197 and PGC\,097958 the gas 
is rotating in a small (1\arcsec\ of diameter) nuclear disk and that {\em 
PA$_{\rm g}$} is nearly aligned to {\em PA$_{\rm s}$}. As a consequence, 
we added these two galaxies to the subsample of objects with no kinematic 
misalignments. In PGC\,047197, we are considering the {\em PA$_{\rm s}$} 
measured within 4\arcsec\ from the center, since the 
kinematics is quite complex outwards (see Fig.~\ref{fig:SATsmaps}). 
The gas in PGC\,046860 and PGC\,047202 does not exhibit a regular rotation
pattern, thus it is not possible to define a {\em PA$_{\rm g}$}.

PGC\,047177 presents discrepant velocities in the inner 2\arcsec\ with 
respect to the outer gas. This did not influence the analysis since the 
inner gas is nearly ``counter-rotating'' with respect to the outer gas 
with {\em PA$_{\rm g}$} nearly aligned with {\em PA$_{\rm s}$}. 
{\em PA$_{\rm g}$} of PGC\,047273, PGC047590, PGC\,049940 and PGC\,065588 was 
straightforwardly derived since the gas is regularly rotating. On the other
hand, the rotation in PGC\,073000 is slightly disturbed by the presence of a 
thick filament. Nevertheless, it was possible to robustly derive the {\em 
PA$_{\rm g}$}. In PGC\,015524, the filaments are more elongated and heavily 
affect the rotation of the gas. To address this issue and obtain a reliable 
{\em PA$_{\rm g}$}, only the central spatial region (within 8\arcsec\ of radius) 
was considered in the procedure and the filaments were discarded. 

Overall, five satellites and four BCGs are characterized by rotating gas. The 
orientations of stellar and gaseous rotations are aligned with respect 
to each other in 3/5 satellites (PGC\,047177, PGC\,047197, PGC\,097958) and 
their {\em PA$_{\rm g}$} and {\em PA$_{\rm s}$} are aligned to the major
axis of isophotes, except for PGC\,047197 (with a difference between 
{\em PA$_{\rm g}$} and $PA$ of the isophotal major axis $\Delta${\em 
PA$_{\rm gp}$} $\sim90^{\circ}$, see Table~\ref{tab:sample}). On the 
contrary, the gas rotation in PGC\,047273 and PGC\,047590 show misalignments 
of the same amount with respect to both the stellar rotation and isophotal 
major axis. 

Concerning the BCGs, the ones with more extended and misaligned gas are those 
with filaments (PGC\,015524 and PGC\,073000). For PGC\,015524, {\em PA$_{\rm g}$} 
is strongly misaligned with respect to {\em PA$_{\rm s}$} ($\Delta${\em PA$_{\rm gs}$} 
$\sim90^{\circ}$), while it is only slightly misaligned 
with respect to the $PA$ of the major axis if one considers the clockwise 
$\Delta${\em PA$_{\rm gp}$}. In PGC\,073000 the stellar and gas kinematics are 
slightly misaligned considering the errors, while they are strongly misaligned 
with respect to the isophotal orientation. On the other hand, PGC\,065588 and 
PGC\,049940 show small and no misalignments, respectively, both between 
{\em PA$_{\rm g}$} and {\em PA$_{\rm s}$}, and {\em PA$_{\rm g}$} and 
isophotal $PA$. 


\begin{figure*}[!t]
\includegraphics[width=\textwidth]{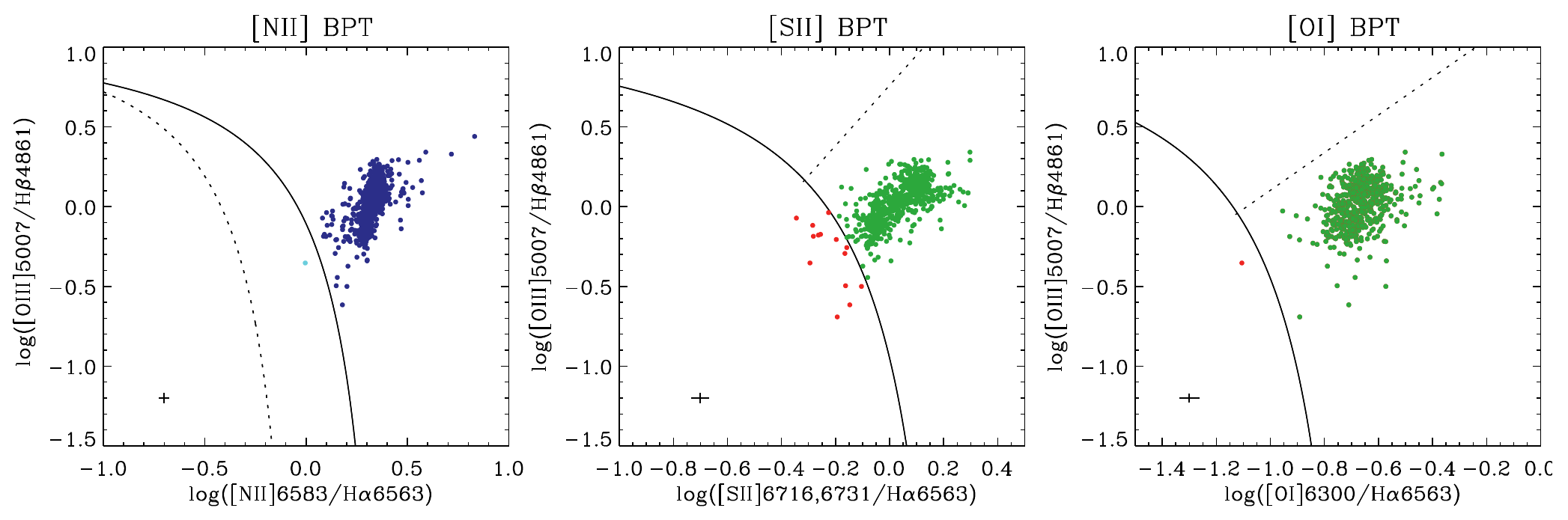} 
\includegraphics[width=\textwidth]{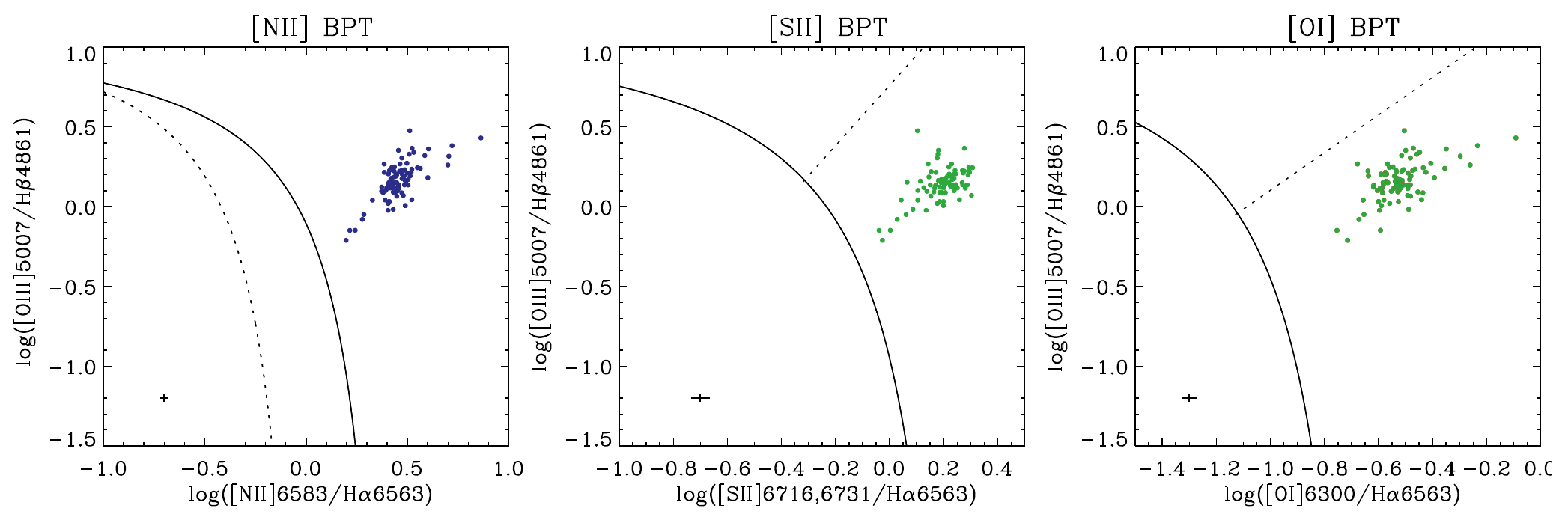} 
\includegraphics[width=\textwidth]{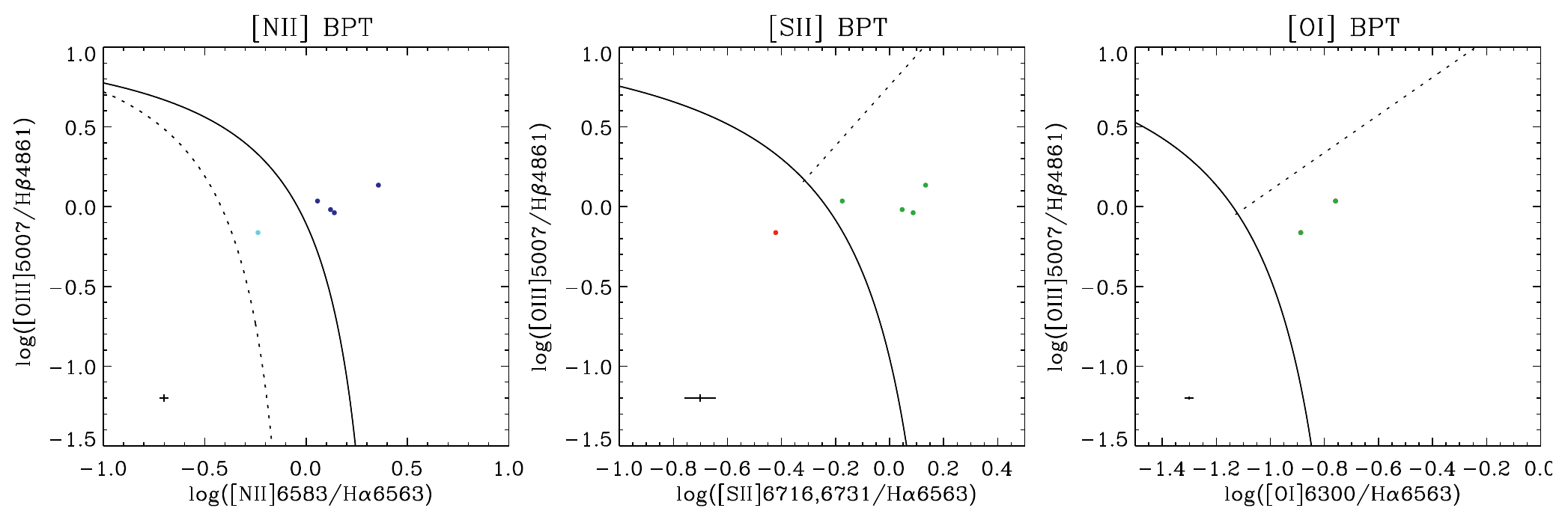} 
\caption{BPT diagrams of PGC\,073000. {\em Left to right}: \niis-BPT, \siis-BPT 
and \ois-BPT diagrams. {\em Up to down}: BPT diagrams of the first, blueshifted 
and redshifted components. The solid curves define the theoretical upper 
bound for pure star formation \citep{Kewley2001}, the dashed curve in \niis-BPT 
defines the upper bound for star-formation dominated bins \citep{Kauffmann2003} 
and the dashed lines in \siis-BPT and \ois-BPT divide Seyfert galaxies from 
LINERs \citep{Kewley2006}. In the bottom-left corner of each diagram the median 
$\pm$3$\sigma$ errors are reported.
Seyfert-type ionization is marked in blue, LI(N)ER regions in the \siis-BPTs 
and \ois-BPTs are displayed in green, SF-dominated bins in red, while composite 
regions between \citet{Kewley2001} and \citet{Kauffmann2003} curves in \niis-BPTs 
are marked in light blue.}
\label{fig:extBPT2}
\end{figure*}

\begin{figure*}
\includegraphics[width=\textwidth]{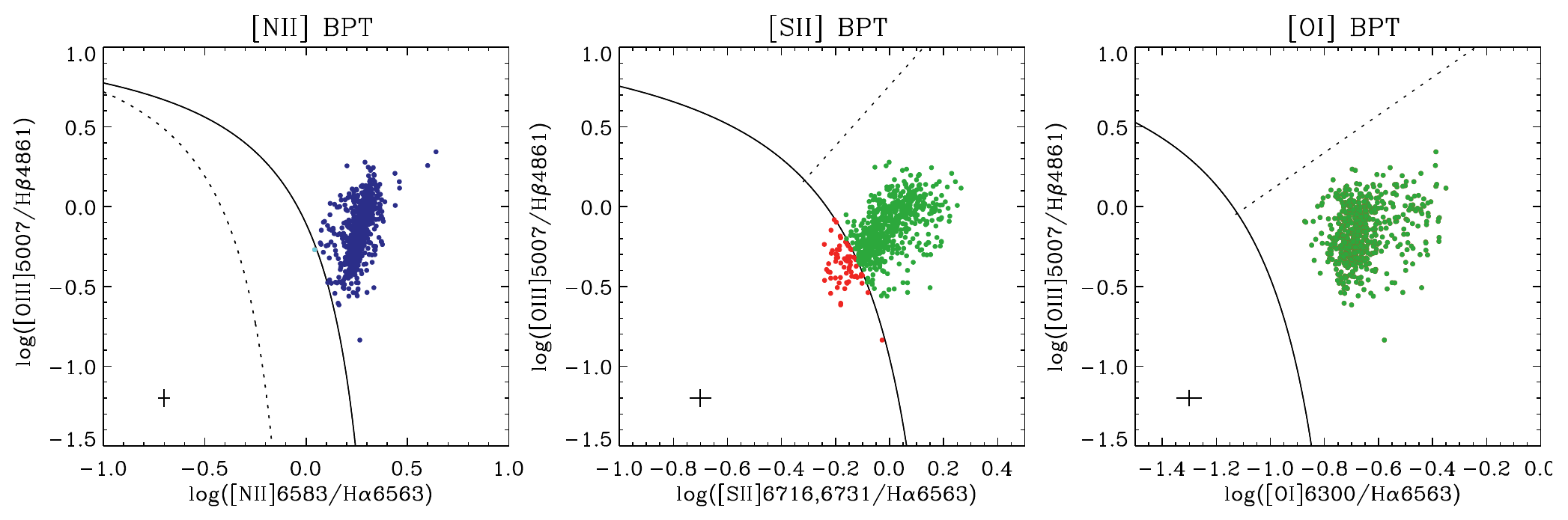} 
\includegraphics[width=\textwidth]{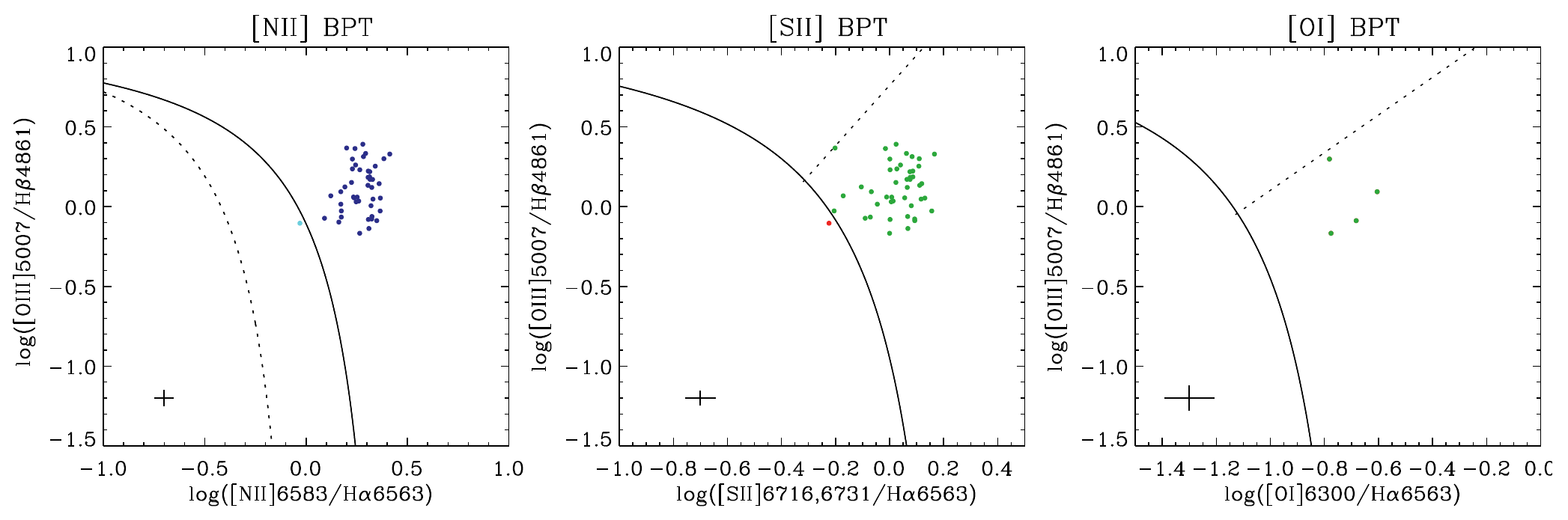} 
\includegraphics[width=\textwidth]{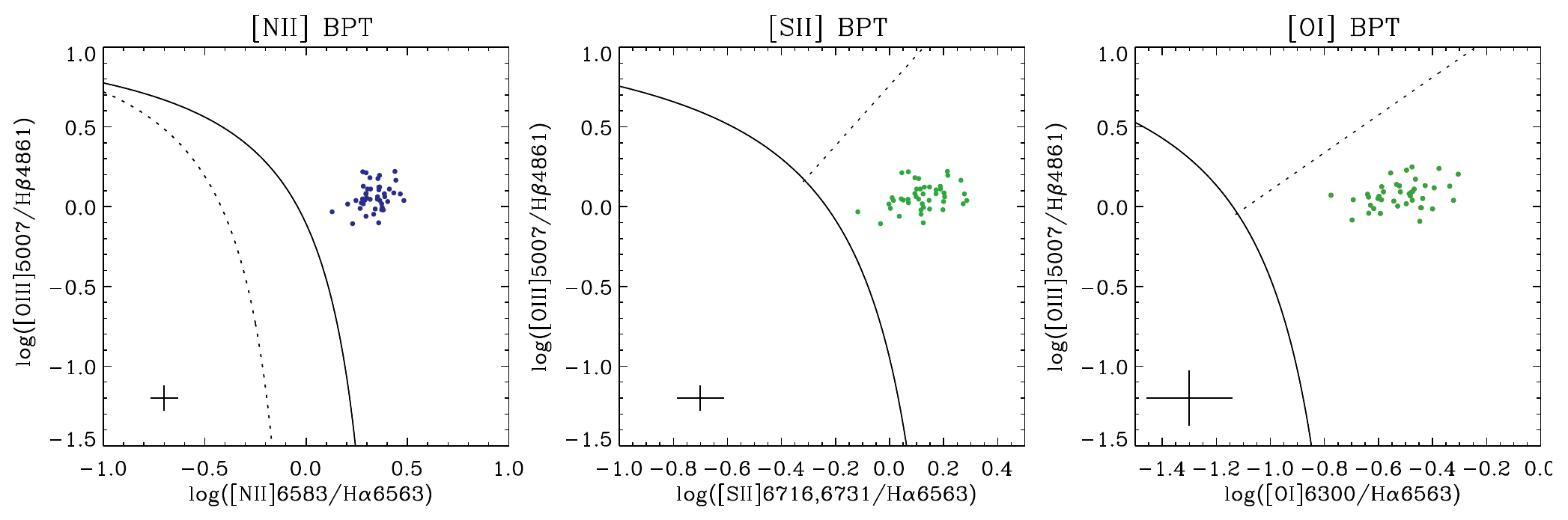} 
\includegraphics[width=\textwidth, trim= 0cm 2mm 0cm 0cm, clip]{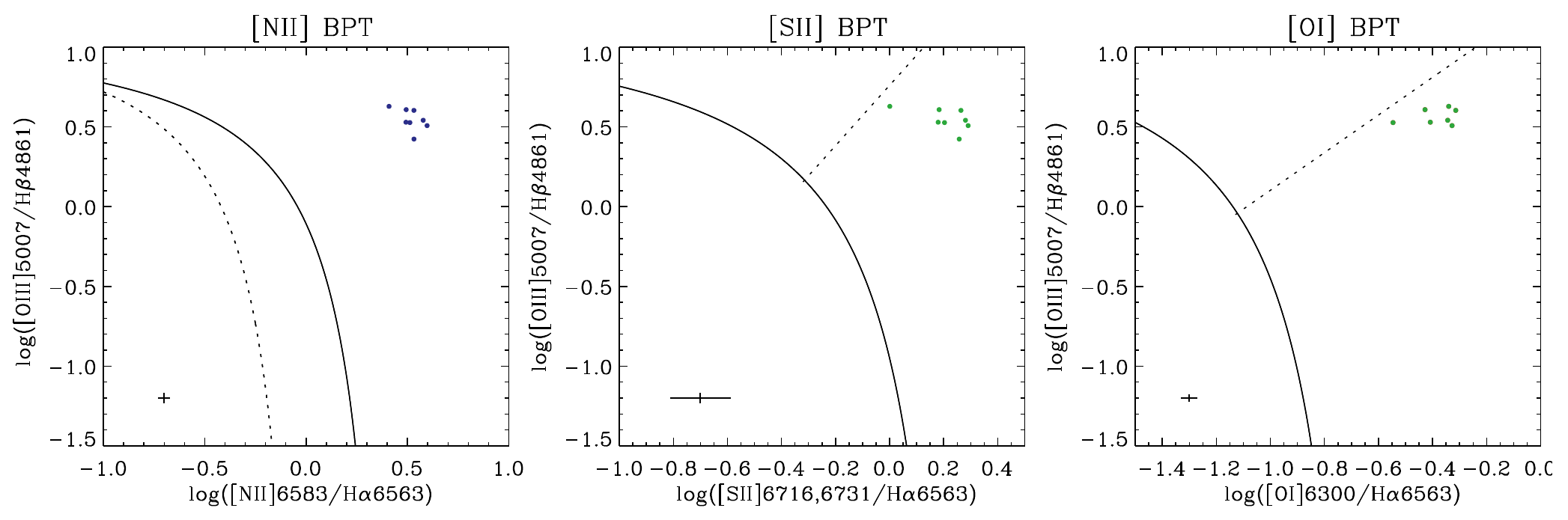} 
\caption{As in Fig.~\ref{fig:extBPT2}. {\em Up to down}: BPT diagrams of the first and 
blueshifted component of PGC\,015524, and first components of PGC\,047177 and PGC\,049940.}
\label{fig:extBPT1}
\end{figure*}


As can be seen, the sample is heterogeneous and too small to draw  
conclusions about the incidence of misalignments in this kind of objects.
Moreover, BCGs and massive satellites should not be gathered in the same 
sample given that they are characterized by different (even if similar 
in some aspects) formation histories. Nevertheless, it has to be pointed 
out that the orientations of stellar and gaseous rotations are aligned 
for 60\% of satellites and 25\% of BCGs. Only 20\% of satellites present 
{\em PA$_{\rm g}$} significantly misaligned with respect to the isophotal 
major axis, while this is true for 75\% of BCGs. 

We used the values of the specific angular momentum ($\lambda_r$), which will 
be presented in a future paper (Krajnovic et al. in prep), to correlate the 
gas properties by splitting the M3G sample in fast and slow rotators. In the 
M3G sample there are 20 slow rotators and 5 fast rotators (all among satellites). 
The sample of fast rotators is not so large to derive statistically significant 
information (due to the mass selection of the sample), but overall the gas was 
detected in 80\% of fast rotators and 35\% of slow rotators. These fractions are 
very similar to the ones found in the MASSIVE survey (80\% and 28\%, respectively, 
\citealt{Pandya2017}). The fraction of slow rotators with gas is larger with respect 
to the one in the MASSIVE sample probably because the M3G sample comprises several 
BCGs that could easily have accreted gas during their frequent mergers (more than 
large ellipticals, \citealt{Edwards2012}; \citealt{Jimmy2013}) or cooling episodes 
when the halo gas can be accreted to the centre of the galaxy \citep{Gaspari2018, 
McDonald2018, Olivares2019}.

In the subsample composed of galaxies with rotating gas, there are 3 fast rotators 
(PGC\,047177, PGC\,047273 and PGC\,097958) and 6 slow rotators (other objects). Slow 
rotators have stronger misalignments ($4/6 \sim 67$\% versus $2/2 = 100$\% misalignments 
in MASSIVE), while fast rotators are predominantly aligned ($1/3 = 33$\% versus $3/4 =75$\% 
misalignments in MASSIVE). 
  
 
\section{Ionisation mechanism}
\label{sec:BPT} 

In order to get a first insight into the ionisation mechanism, we analysed 
the emission line flux ratios of all velocity components separately. We built 
BPT diagrams \citep{Baldwin1981, Veilleux1987} by considering \niis/H$\alpha$ 
(\niis-BPT), \siis/H$\alpha$ (\siis-BPT) and \ois/H$\alpha$ (\ois-BPT) flux 
ratios of 3 BCGs (PGC\,015524, PGC\,049940, PGC\,073000) and one SSC satellite 
(PGC\,047177) for which both H$\beta$ and \oiiig\ were detected. For PGC\,047202, 
PGC\,047273, PGC\,047590 and PGC\,065588 it was possible to analyse \niis/H$\alpha$\ 
ratios only. Finally, only the \niis\ line was clearly detected in PGC\,046860, 
PGC\,047197 and PGC\,097958 and this prevented us from studying their ionisation 
mechanism. 

The ionisation in massive ETGs on kpc scales is commonly due to photoionization 
from hot underlying evolved stars and shocks, while in the nuclear region also 
low-luminosity AGN may contribute \citep{Pandya2017}. Thanks to spatially resolved 
spectroscopy, extended LIER emissions have been discovered (e.g., \citealt{Belfiore2016}) 
and are ascrible to ionizing photons from post-Asymptotic Giant Branch (pAGB) stars
(\citealt{Sarzi2006}; \citealt{Johansson2016}). 

The BPT diagrams are shown in Fig.~\ref{fig:extBPT2}, and 
Fig.~\ref{fig:extBPT1}. Probably due to extinction, in  PGC\,047177 and PGC\,049940 
the second component was not detected in H$\beta$ and \oiiis\ line profiles, and 
we built the BPT diagrams only for their first components. The flux line ratios of 
these two galaxies are enclosed in the LINER region. Overall, the gas in PGC\,049940 
extends up to 2\arcsec\ from the center but only the nuclear bins (within 0\farcs8 of 
radius that corresponds to 0.6 kpc) are considered in the BPTs because H$\beta$ 
was not detected in the outer ones while the bins of PGC\,047177 considered in the 
BPT are enclosed within 2\arcsec\ of radius (that corresponds to 1.9 kpc). Notably, 
PGC\,049940 shows higher \oiiis/H$\beta$ and \niis/H$\alpha$ ratios but nearly the 
same \siis/H$\alpha$ ratio as PGC\,047177. 

PGC\,073000 and PGC\,015524 are characterized by spatially extended emission up to about 
10\arcsec\ and 5\arcsec, corresponding to 6.8 and 5.0 kpc, respectively, plus filaments
elongated until 18\arcsec. The BPT diagrams of the first component of both galaxies show 
predominately LINER line ratio with little contamination from star formation (similarly 
to \citealt{Loubser2013} and \citealt{McDonald2012} for PGC\,015524) in the \siis-BPT, even 
if this is not visible in the \niis-BPT diagram. The absence of a significant gradient in 
\niis/H$\alpha$ ratio and the low ratio between \oiiis/H$\beta$ allow us to conclude that 
the gas ionisation is primarily not due to the central AGN. All of the line ratios of 
red/blueshifted components fall in the LINER region (see Fig.~\ref{fig:extBPT2} and
 \ref{fig:extBPT1}). 

Generally speaking, within the nuclear region the LINER line ratios could be related 
to shocks produced by the outflows (presumably observed in our galaxies) along with 
emission from the AGN accretion disk \citep{Molina2018} or underlying old stellar 
population. PGC\,049940 can be classified as type-1 LINER\citep{HermosaMunoz2020} 
due to the presence of a broad H$\alpha$ emission in the nucleus coming from the 
BLR. In this case we concluded that the AGN contribution to ionisation is strong 
in the nucleus. On the contrary, PGC\,015524, PGC\,047177 and PGC\,073000 can be 
classified as type-2 LINER \citep{Cazzoli2018} due to the absence of a BLR 
H$\alpha$ emission. It has to be pointed out that the broad H$\alpha$ emission
may not be detected in ground-based data especially in presence of additional 
components that may hide it and hamper the spectroscopic classification. This 
is not the case of PGC\,047177 that presents, even in the nucleus, a weak second 
component for each line, and PGC\,073000 with line profile asymmetries unequivocally 
ascribable to red/blueshifted additional components. The superb $S/N$ in these 
spectra allows to clearly distinguish different components in most of blended line
profiles. In these two galaxies both shock fronts and evolved stellar populations 
may provide their contribution to the ionisation. The fit of the nuclear spectra 
of PGC\,015524 is more uncertain since H$\alpha$ and \niis\ lines are strongly blended. 
Nevertheless, the nuclear line profiles are clearly bent towards shorter wavelengths 
and employing a blueshifted component for each line allowed to successfully fit them. 
This component is broader in the nucleus (about 480 \kms) than outside (about 250 \kms), 
but still not so broad to arise from the BLR and so this emission-line components are 
likely to be excited by radiative shocks. 

The LIER (out of the nucleus) line ratios may arise from a mixed contribution of 
evolved stellar populations \citep{Hsieh2017} that are quite common in these kind 
of objects, and shocks due to galactic-scale outflows. 

Concerning the spatially extended first components of PGC\,015524, 
PGC\,047177, and PGC\,073000, the ionisation mechanism could be ascribable 
to a mixed contribution from the underlying evolved stellar populations 
and shallow shocks. \citep{Byler2019} predicted the LIER-like emission 
in ETGs from hot post-AGB stars based on a self-consistent stellar and 
photoionization models. Looking at their \siis\ and \ois- BPT diagrams, 
LIER predictions for old stellar populations (older than 10 Gyr) fall 
slightly below (0.5 dex) with respect to the line ratios of PGC\,015524, 
PGC\,047177, and PGC\,073000. 

The velocity dispersion is a suitable shock diagnostic \citep{Ho2014} for outer 
regions where there is no AGN contribution \citep{dAgostino2019}. Indeed, the 
gas that was ionized by shocks presents both high flux ratios and high velocity 
dispersion between 150 and 500 \kms. This is exactly the case of red/blueshifted 
components revealed in our galaxies, especially for PGC\,047177 and PGC\,049940. 
Consistently with the presence of shocks, \oi\ and \ni\ emission were detected. 
This gas is often situated in extended and partially ionized regions produced by the hard radiation field of the shock. 
In PGC\,015524 and PGC\,073000, \ois\ and \nis\ show the same kinematics and similar 
spatial distribution of the ionised-gas, even if they are less spatially extended 
outward (especially the \nis). In particular, \ois\ and \nis\ have a filamentary 
distribution in PGC\,015524, while in PGC\,073000 they are more concentrated in the 
center and in the west direction in correspondence to the X-ray arm. 


\begin{figure*}[!t]
\includegraphics[scale=0.2]{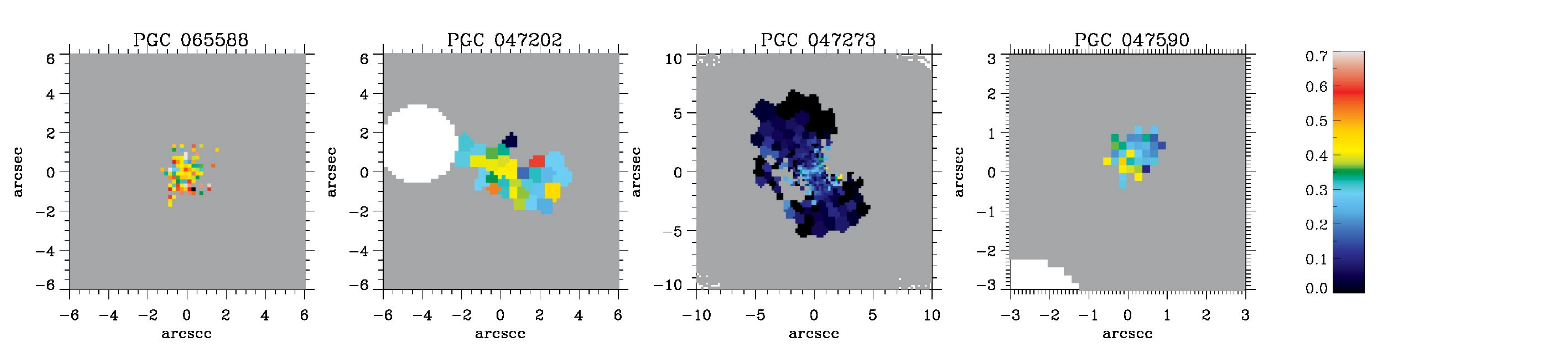} 
\caption{{\em Left to right}: $\log$ (\niis/H$\alpha$) spatially resolved maps  
of PGC\,065588, PGC\,047202, PGC\,047273 and PGC047590. Their spatial 
extent is the same of H$\alpha$ that is less extended than \niis\ in 
all the galaxies.}
\label{fig:NIIHA}
\end{figure*}

For PGC\,047202, PGC\,047273, PGC\,047590 and PGC\,065588 that are characterized 
by too weak \Hbs\ or no \oiiis\ emission, the $\log$ \niis/H$\alpha$\ ratios 
are reported in Fig.~\ref{fig:NIIHA}. This ratio is a good indicator of gas 
excitation state caused by an AGN. Indeed, it saturates at high metallicity 
(\citealt{Kewley2006}) and, as a consequence, ratios larger than $\sim0.6$ 
points to an AGN contribution (rather than star formation) to the excitation \citep{Osterbrock2006,Hamer2016} even if it does not have a decisive discriminating 
power. Moreover, star formation is characterized by softer radiation field with 
respect to AGN, which has a harder radiation producing a higher ionization state 
and higher $\log$ \niis/H$\alpha$\ ratios.

For PGC\,047273 and PGC\,047590, the ratio varies within the resolved maps  
(up to about 0.3 and 0.4, respectively, see Fig.~\ref{fig:NIIHA}) but it 
definitely remains under 0.6, ruling out AGN/LINER contamination. The 
same considerations can be applied to the BCG of SSC (PGC\,047202), even 
if there is a stronger spatial variation. PGC\,065588 is characterized by 
higher ratios (on average above 0.4 and up to 0.8 for some bins) even if 
there is not a clear radial variation. In case of contamination by an AGN 
one would expect high ratios more localized in the center while the spatial 
distribution of ratios points to a composite LINER and star formation 
contribution. Nevertheless, $\log$ \niis/H$\alpha$\ values are not  
sufficient to determine the ionisation mechanism since a number 
of processes could play a role, like AGN, star formation, shocks from 
internal or external sources (faster shocks easily result in higher line 
ratios), photoionization by hot gas, and collisional heating. 


\begin{figure*}
\includegraphics[scale=0.18]{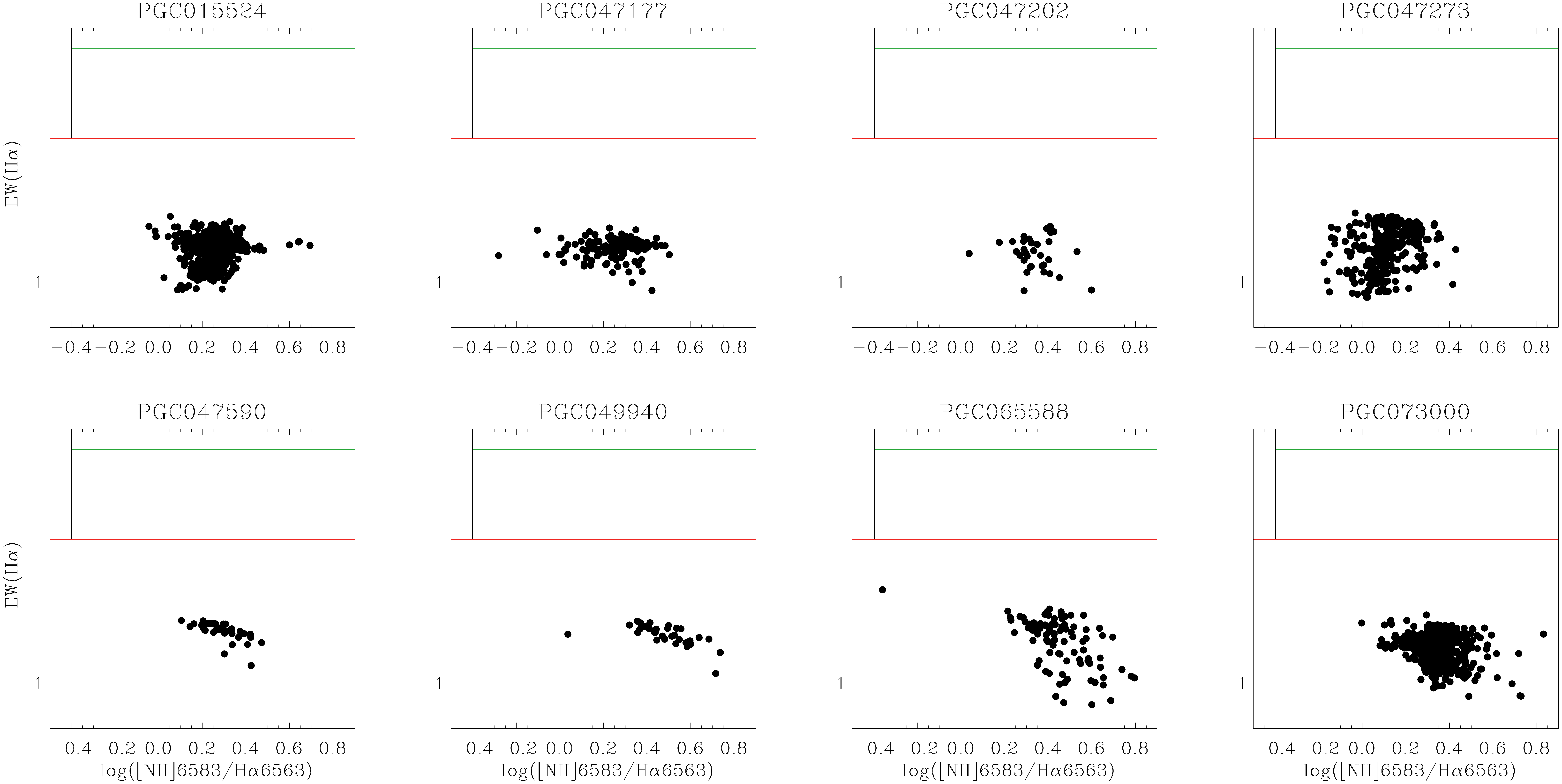} 
\caption{{\em Left to right}: WHAN diagram. 
The lines represents the division between gas ionised by AGN/star formation 
and ``retired'' galaxies (above and below the red line, respectively), the optimal 
transposition of the BPT-based division \citep{Grazyna2006} between ionisation 
mechanisms driven by star formation and AGN (to the left and right of the black line,
respectively) and of the \citet{Kewley2006} separation between Seyferts and LINERs
(above and below the green line, respectively). The name of the galaxies is reported 
above each panel.}
\label{fig:EWHA}
\end{figure*}

For those galaxies with both H$\alpha$ and \niis\ lines detected we built 
the WHAN diagram (\citealt{CidFernandes2011}; \citealt{Sanchez2020_2}) 
that relates the equivalent width of H$\alpha$ and the \niis/H$\alpha$ ratio. 
For those galaxies with emission lines with more than one component, we considered 
for this analysis the first one. The WHAN diagram provides more information on the 
ionisation mechanism for those galaxies for which it is not possible to built 
the BPT diagram. Moreover, it allows to separate weakly active galactic nuclei 
and passive galaxies ionized by the underlying evolved stellar population that 
fall both on the LI(N)ER region of the BPT diagram. As can be noticed in 
Fig.~\ref{fig:EWHA}, for all galaxies the data points are enclosed in the region 
of the diagram corresponding to ``retired'' galaxies, i.e. objects that stopped 
forming stars and with gas ionized by hot low-mass evolved stars. This points 
out an important contribution to the ionisation from the underlying evolved 
stellar population.


The determination of the relative contributions to ionisation due to shocks, 
underlying stellar population or AGN in LINERs requires the employment of 
photoionization models in order to interpret the emission line ratios. 
Moreover, an analysis of UV and/or IR spectroscopic data would help (see 
\citealt{Kewley2019} for a review). This is beyond the scope of this paper. 


\section{The origin of the gas}
\label{sec:discussion} 

The M3G sample was built to select the most massive galaxies in 
the local Universe without putting constraints on gas content and, 
as a consequence, the sub-sample of galaxies with gas is extremely 
heterogeneous but illustrative at the highest galaxy 
masses. In the following, we exploit the previously derived gas 
properties to infer hints concerning the origin of the gas and give 
an explanation to the observed distribution and kinematics. Moreover, 
the comparison between the orientations of stellar and gaseous rotations 
(see Section~\ref{sec:orientat}) along with the specific angular momentum 
information provide us with crucial clues on the origin of the gas given 
that slow rotators are more likely to show misalignments because of past 
merging activity.


\subsection{Galaxies with extended gas and filaments}

PGC\,015524 and PGC\,073000 show extended emission with filamentary structures. 

In PGC\,015524, the gas stretches in filaments with similar distribution 
suggesting a common origin, even if the filaments are less extended for \nis\ 
and \ois. \citet{McDonald2012} carried out a long-slit spectroscopic analysis 
along the filaments and proposed that the gas clouds would have reached the 
cluster center through gravitational free fall. Contrary to this scenario, 
changes in the sign of the line-of-sight velocity are clearly visible in 
our spatially resolved maps of H$\alpha$, \niis, and \siis, rather pointing 
to the so called ``rising bubble''mechanism (e.g., \citealt{Reynolds2005}). 
Nevertheless, there are no clear bubbles which should form the filamentary 
structures \citep{Dunn2006}. Moreover, the galaxy has a cooling flow 
\citep{Edwards2007} visible in X-ray gas emission and \citet{Dunn2006} 
stated that there should be some heating mechanism at work based on the 
cooling times as confirmed by \citet{McDonald2018}. Using deep Chandra 
observations, \citet{Hogan2017} found that the extended H$\alpha$ emission 
is associated to the cooler X-ray component in a two-temperature thermal 
model of the ICM. This suggests that the gas originates from a direct 
cooling of the hot phase. 
From our analysis, we found that there is a strong misalignment (91\degr) 
between the gas and stellar kinematics, supporting an external accretion 
of the gas. 

In PGC\,073000, ionized-gas, \ois\ and \nis\ are kinematically aligned 
and share similar distributions pointing to a common origin. Moreover 
all the line profiles are characterised by three gas components with 
different velocities (for an example see Fig.\ref{fig:lines}) suggesting 
that ionized-gas, \ois\ and \nis\ are subjected to the same processes 
(e.g., outflows) within the galaxy. All spatially resolved maps show a 
second flux peak, in addition to the central one, in the west direction 
in correspondence to an X-ray arm \citep{Choi2004}. 
The gas and stellar rotations are nearly aligned with respect to each other 
while they are not with respect to isophotal $PA$ (117\degr). The gas seems 
to be associated with the stellar component that rotates around the major 
axis (prolate-like). In such a configuration, the gas could 
have been acquired by chance along that axis, or be associated with the stellar 
component from the formation of the galaxy. The fact that the velocity map is 
actually quite disturbed suggests that this gas could indeed be infalling and 
settling into this principal plane due to the gravitational torques. The thick 
filament, that is visible in the \niis, \siis, and H$\alpha$ maps only, presents 
no changes in the sign of the velocity, pointing to a later stage of accretion 
with respect to PGC\,015524.
Furthermore, this galaxy has a cooling flow in action \citep{Edwards2007} 
and the filament may be formed by gravitational free fall of cold gas. 
Finally, PGC\,073000 hosts radio emission towards the north direction 
\citep{McDonald2010} that may be related to some perturbations visible 
in the gas distribution. 


\subsection{Galaxies with centrally concentrated gas}

In three BCGs (PGC\,047202, PGC\,065588 and PGC\,049940) and four SSC satellites 
(PGC\,046860, PGC\,047197, PGC\,047590 and PGC\,097958) the gas is confined 
in the central region in a compact ionised-gas disks in rotation, except 
for PGC\,046860 and PGC\,047202 in which the gas is not rotating. 

The rotation of the gaseous disk in PGC\,049940 is aligned to the stellar one. 
PGC\,047197 and PGC\,097958 are ascribable to the same category, even if the 
putative alignment is more uncertain given that this information was derived
from visual inspection of few bins. This small amount of gas contained in 
objects with oblate-like regular rotation (PGC\,046860, PGC\,047197 and PGC\,049940)
may originate from internal stellar gas loss, even if simulations showed that 
it could also come from mergers \citep{Voort2015}. In particular, this could be 
the case of PGC\,047197 which shows prolate-like rotation \citep{krajnovic2018}.

On the contrary, in PGC\,065588 and PGC\,047590 the rotation of the gas 
is not aligned to the one of stars suggesting a possible external origin 
of the gas. On the other hand, an internal origin is still plausible given 
that they are characterised by prolate-like stellar rotation \citep{krajnovic2018}. 
Indeed, the gas lost from the stars, that belong to various orbital families, can 
end up in a configuration different from the ones of the stars through several 
cooling cycles and its present configuration could just be a consequence 
of it being released by stars on different orbits. This could also be the case 
of PGC\,047197.


\subsection{Peculiar gas distributions and kinematics}

PGC\,047177 and PGC\,047273 present peculiar features in the gas distribution 
and kinematics so that it is more difficult to speculate on the origin of their 
gas.   

In PGC\,047177, {\em PA$_{\rm g}$} and {\em PA$_{\rm s}$} are aligned 
with respect to each other and their rotation axis are aligned to the 
isophotal major axis. Nevertheless, the H$\alpha$ distribution is a bit 
filamentary and patchy as the \niis\ and \siis\ ones that show a similar 
pattern even if they are more compact. More peculiar, the velocity field 
shows a change of the sign, of the order of 50 \kms, in the inner spatial 
region. A counter-rotation is not visible in the stellar velocity map and, 
unfortunately, there are no high-resolution HST images. As a matter of fact, 
in a regular and fast rotator as PGC\,047177 this is quite rare. Nevertheless, 
the presence of nuclear dust (see Sec.~\ref{sec:reddening}) could be interpreted 
as the smoking gun for a gas disk, with the same spatial scale, that was recently 
acquired in multiple phase accretion process and settled in a different orientation. 
This interpretation has to be taken with caution since the evidence is marginal. 

PGC\,047273 is characterized by an oblate-like regular rotation but presents 
a large amount of ionized-gas with a rotation axis that is strongly misaligned 
with respect to the stellar one. The presence of misalignment between stars and 
gas points to an external origin of the gas. After the accretion, 
the gas settled in a sort of polar disk which is, however, not exactly along the 
galaxy minor axis. This could suggest that PGC\,047273 is a triaxial galaxy observed 
under a special orientation. In a triaxial galaxy that would be a stable configuration, 
but in an axisymmetric one it would be less stable. 
Fig.~\ref{fig:047273} shows all the detected emission-lines for this galaxy, 
some with more extended coverage of the gas distribution. From these lines 
it can be noticed that in the outskirts the disk is visibly ``precessing'' 
and bending towards the major axis of the galaxy. This suggests that the 
galaxy might be oblate axisymmetric and the accreted gas disk 
is in the process of setting in the principal plane of the galaxy. 


\section{Conclusions and future work}
\label{sec:conclusions}

We analysed the gas content of massive (M $> 10^{12}$ \msun) ETGs in the 
M3G survey by exploiting MUSE data. Ionised-gas was detected in 11 galaxies 
along with \ois\ and \nis\ in 4 of them. Our main results can be summarised 
as follows:

\begin{itemize}

\item[$\bullet$] In the M3G sample, $\sim$28\% of BCGs and $\sim64$\% of 
massive satellites of the SSC contain ionised-gas. Among them, \ois\ and 
\nis\ were detected in three BCGs and one satellite. These four objects 
present multi-component line profiles, with red/blueshifted and broad 
components. 

\item[$\bullet$] The gas distributions and kinematics are extremely 
heterogeneous in these objects. The gas is centrally concentrated in 
almost all objects, except for two BCGs (PGC\,015524 and PGC\,073000) 
that show filaments and two massive satellites with extended 
emission (PGC\,047177 and PGC\,047273). The peak of the gas emission 
corresponds to the photometric center for all galaxies (except for 
PGC\,015524 and PGC\,073000 with perturbed gas distribution in the nuclear
region).

\item[$\bullet$] Dust was probed in such old and massive objects by analysing 
the extinction with Balmer ratios. We revealed mean $E(B-V)$ of 0.2-0.3 values 
with few peaks up to 0.7 corresponding to optically thick dust filaments visible 
also in HST images. 

\item[$\bullet$] The orientations of stellar and gaseous rotations are 
aligned for 60\% of satellites and 25\% of BCGs. 75\% of BCGs and 20\% 
of satellites are characterized by a {\em PA$_{\rm g}$} significantly 
misaligned with respect to the isophotal {\em PA}. Overall, the gas was 
detected in 80\% of fast rotators and 35\% of slow rotators, similarly 
to the MASSIVE survey galaxies.

\item[$\bullet$] From the analysis of \niis-BPT and \siis-BPT we conclude that 
PGC\,015524, PGC\,047177, PGC\,049940 and PGC\,073000 are characterised by LI(N)ER 
line ratios, although PGC\,015524 and PGC\,073000 may show possible contamination 
from star formation. We classified PGC\,049940 as type-1 LINER while PGC\,015524, 
PGC\,047177 and PGC\,073000 as type-2 LINER. High velocity dispersions up to 600 
\kms\ were detected for the red/blueshifted components in the nuclear region, 
pointing to the presence of shock excitation. Extended gas in the red/blueshifted 
of PGC\,015524, PGC\,047177, PGC\,049940 and PGC\,073000 are probably mainly ionised 
by the underlying evolved stars (pAGBs) with additional shock contribution, 
given the high \ois/H$\alpha$ and \siis/H$\alpha$ ratios, the presence of \ois\ 
and \nis\ emission and the high velocity dispersion.

\item[$\bullet$] The presence of misalignments between {\em PA$_{\rm g}$} 
and {\em PA$_{\rm s}$} may point to an external origin of the gas for 
PGC\,015524, PGC\,047273, PGC\,047590, PGC\,065588 and PGC\,073000 (three BCGs and 
two SSC satellites). On the other hand, it has to be pointed out that PGC\,047197, 
PGC\,047590, and PGC\,065588 are characterized by prolate-like stellar 
rotation and the gas could have been originated internally and settled in 
different configuration with respect to stars.

\item[$\bullet$] The presence of discrepant velocities with respect to the 
background gas structure in the inner 2\arcsec\ of PGC\,047177 would need 
to be inspected with HST images, that are not present in the archive. The 
puzzling case of the ``precessing'' polar disk of PGC\,047273 would require 
further analysis concerning stellar populations and star formation history 
to better characterize the evolution of this massive galaxy.

\end{itemize}

Future work comprises the determination of the relative contributions to ionisation 
due to shocks, underlying stellar population or AGN in LINERs. It has to be pointed 
out that optical diagnostics and photoionization models are not ideal for separating 
excitation sources if shocks are present, since shocks models are located in the same 
BPT region as the ones of AGNs \citep{Kewley2019}. In this sense, the analysis of IR and/or 
UV spectroscopic data would appreciably help to separate the contributions to ionisation. 
Moreover, the study of stellar populations will provide decisive insights into the 
formation histories of these ETGs, confirming or reconsidering the discussion in 
Section~\ref{sec:discussion} about the origin of the gas.


\begin{acknowledgements}
IP acknowledges the Leibniz Institute For Astrophysics Potsdam for the hospitality 
and financial support while this paper was in progress. IP is supported also by 
Fondazione Ing. Aldo Gini of the University of Padova. We are grateful to Lutz 
Wisotzki, Tanya Urrutia and the MUSE GTO collaboration for the valuable comments.
DK and MdB acknowledge financial support through the grant GZ: KR 4548/2-1 of the 
Deutsche Forschungsgemeinschaft. 
JB acknowledges support by Fundação para a Ciência e a Tecnologia (FCT) through 
the research grants UID/FIS/04434/2019, UIDB/04434/2020, UIDP/04434/2020 and 
through the Investigador FCT Contract No. IF/01654/2014/CP1215/CT0003.
PMW received support through BMBF Verbundforschung (project MUSE-NFM, grant 05A17BAA).
WK was supported by DFG grant Ko 857/33-1. 
This research is based on MUSE data of the MUSE GTO collaboration and  
data from the NASA/IPAC Extragalactic Database (http://ned.ipac.caltech.edu/).
\end{acknowledgements}

%
%


\begin{appendix} 
\section{Maps for the gas kinematics, fluxes, and A/N of the M3G galaxies}
\label{sec:appendixA}

We show the 2D spatially resolved maps of the gaseous fluxes, 
kinematics and $A/N$, extracted in Section~\ref{sec:extraction} and 
described and commented in Section~\ref{sec:detect}. 


\begin{figure*}
\centering
\includegraphics[scale = 0.225, trim= 0cm 7.4cm 0cm 0cm, clip]{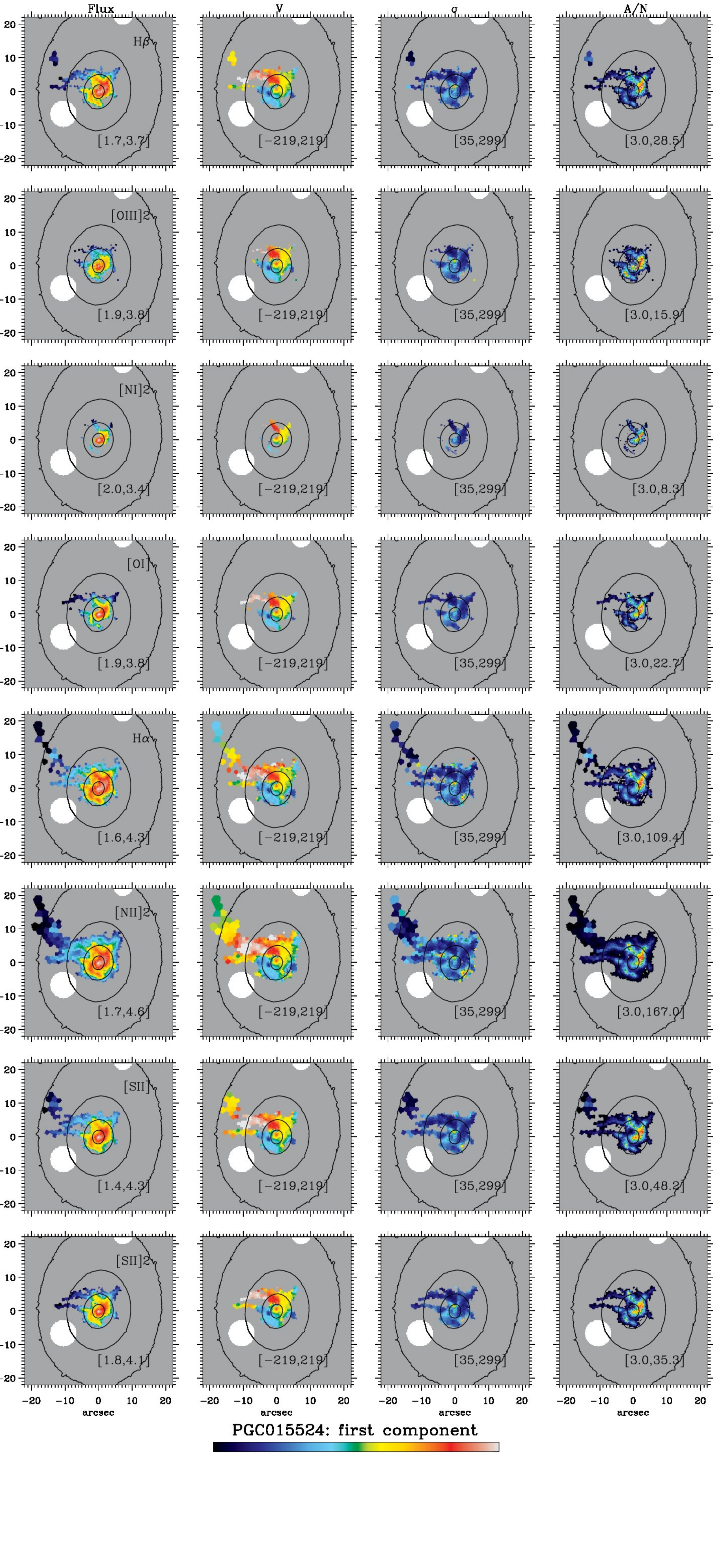} 
\caption{Spatially resolved maps of the first gas velocity component of PGC\,015524. 
{\em Left to right}: flux in $\log$ 10$^{-20}$ erg~s$^{-1}$~cm$^{-2}$~arcsec$^{-2}$, 
velocities and velocity dispersions in \kms, and $A/N$ ratio. The maximum and 
minimum values in the colorbar are reported in each panel between square 
brackets. Black dashed contours are isophotes in steps of one magnitude. {\em Up to down}: 
Spatially resolved maps of H$\beta$, \oiiig\ (\oiiis 2), \nidx\ (\nis 2), \oisx\ (\ois), 
H$\alpha$, \niig\ (\niis 2), \siip\ (\siis) and \siig\ (\siis 2).}
\label{fig:015524_1}
\end{figure*}

\begin{figure*}
\centering
\includegraphics[scale = 0.23, trim= 0cm 7.4cm 0cm 0cm, clip]{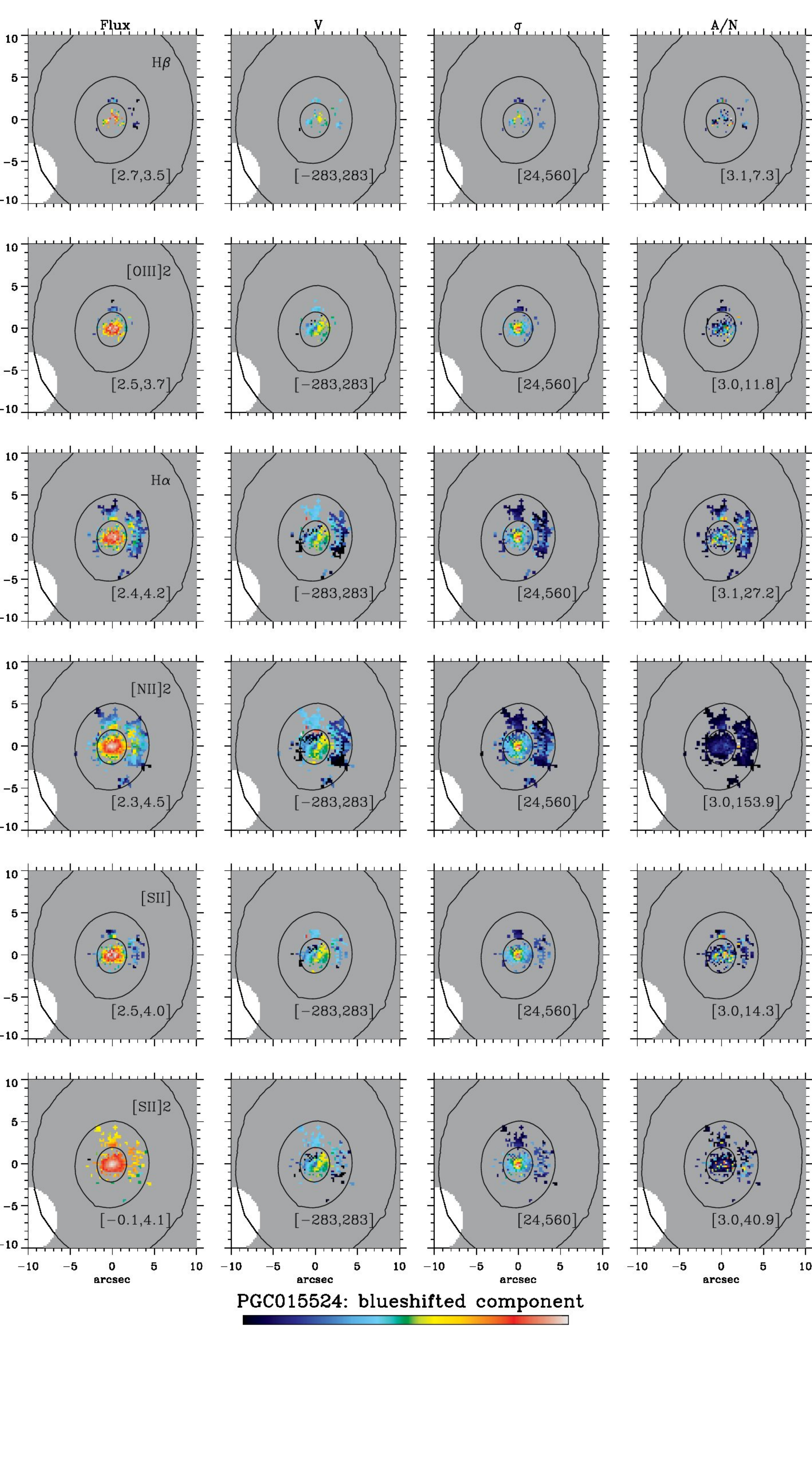} 
\caption{As in Fig.~\ref{fig:015524_1} but for the blueshifted component of PGC\,015524. 
{\em Up to down}: Spatially resolved maps of H$\beta$, \oiiis 2, H$\alpha$, 
\niis 2 and \siis\ and \siis 2.}
\label{fig:015524_2}
\end{figure*}


\begin{figure*}
\centering
\includegraphics[scale = 0.23, trim= 0cm 15cm 0cm 0cm, clip]{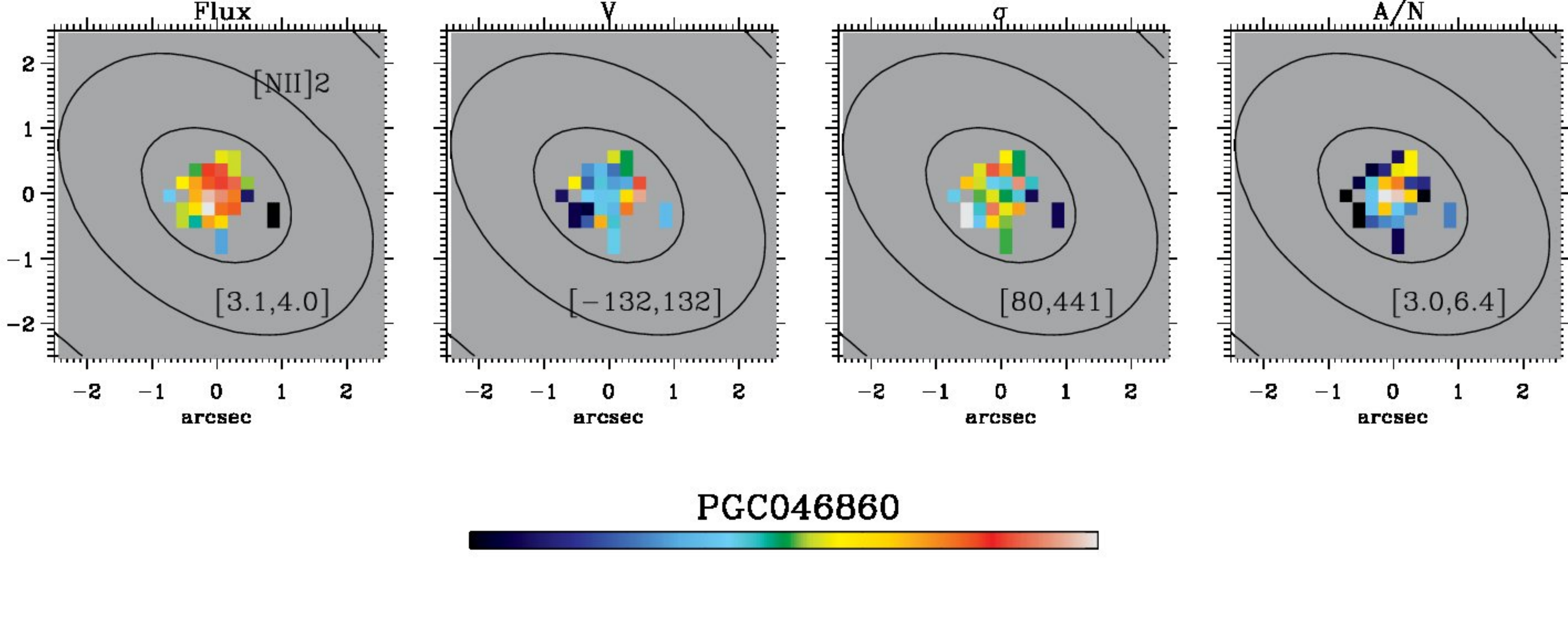} 
\caption{As in Fig.~\ref{fig:015524_1} but for PGC\,046860. 
Spatially resolved map of \niis 2.}
\label{fig:046860}
\end{figure*}


\begin{figure*}
\centering
\includegraphics[scale = 0.23, trim= 0cm 17cm 0cm 0cm, clip]{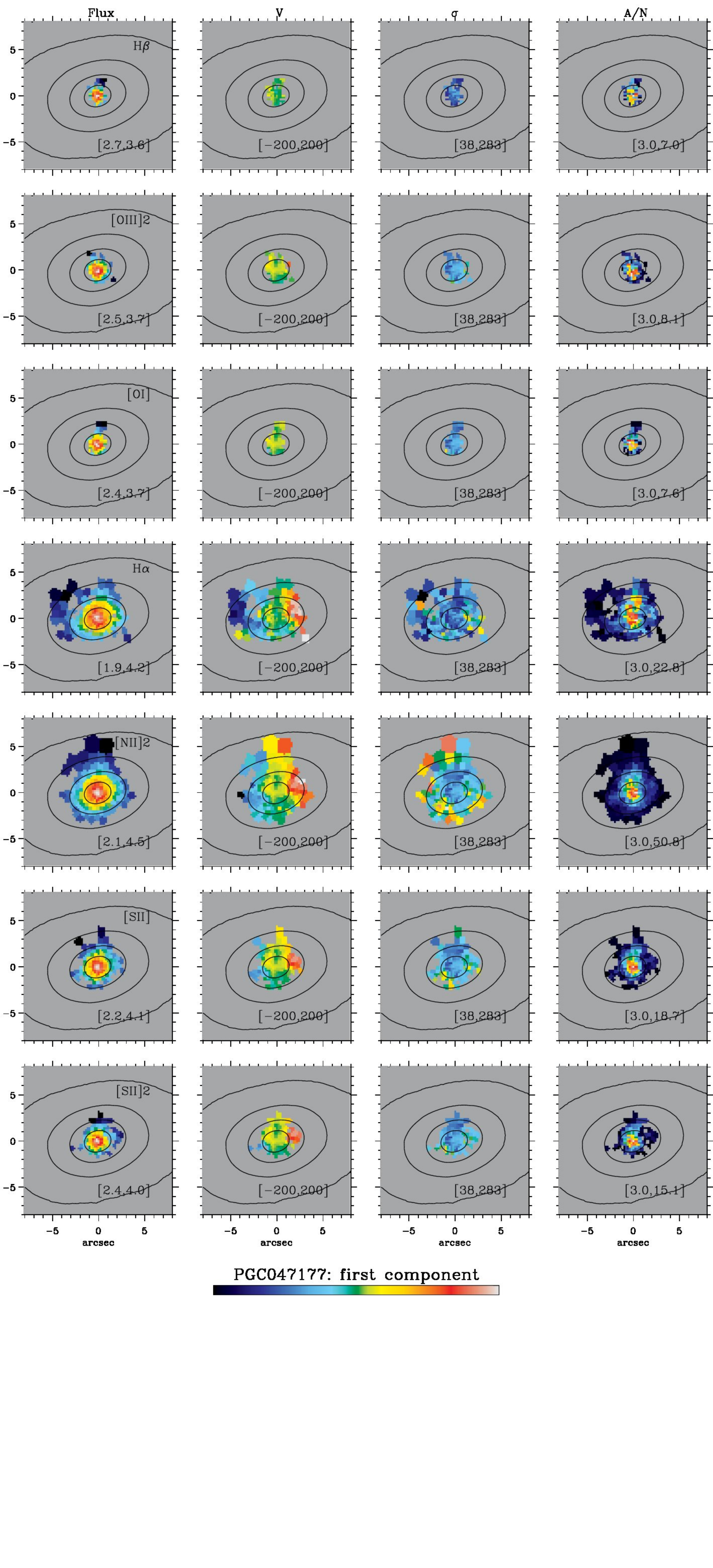} 
\caption{As in Fig~\ref{fig:015524_1}  but for the first gas velocity component of  PGC\,047177. 
{\em Up to down}: Spatially resolved maps of H$\beta$, \oiiis 2, \ois, H$\alpha$, 
\niis 2 and \siis\ and \siis 2.}
\label{fig:047177_1}
\end{figure*}

\begin{figure*}
\centering
\includegraphics[scale = 0.23, trim= 0cm 18cm 0cm 0cm, clip]{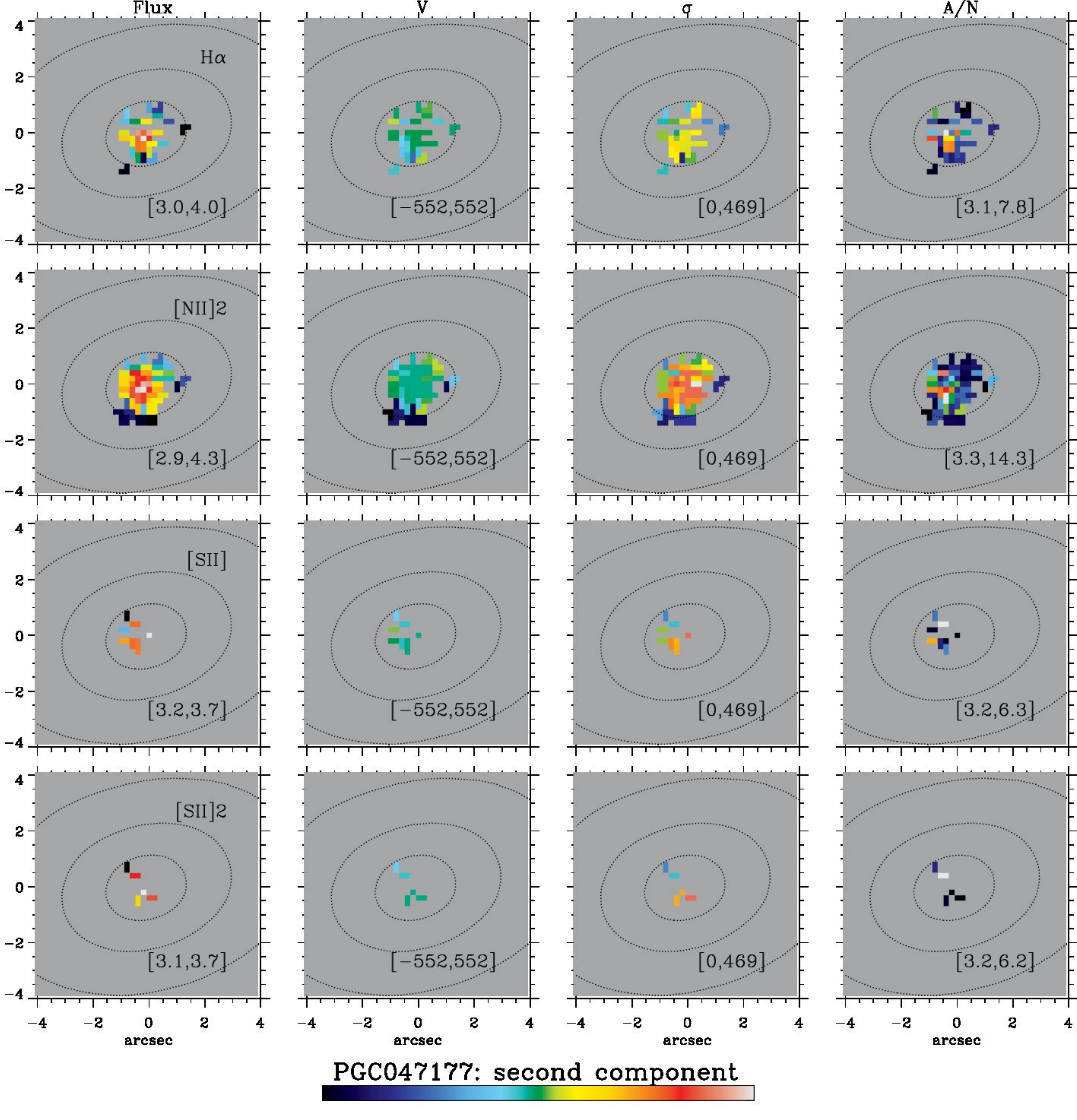} 
\caption{As in Fig~\ref{fig:015524_1}  but for the second velocity component of  PGC\,047177. 
{\em Up to down}: Spatially resolved maps of H$\alpha$, \niis 2 and \siis\ and \siis 2.}
\label{fig:047177_2}
\end{figure*}


\begin{figure*}
\centering
\includegraphics[scale = 0.23, trim= 0cm 16cm 0cm 0cm, clip]{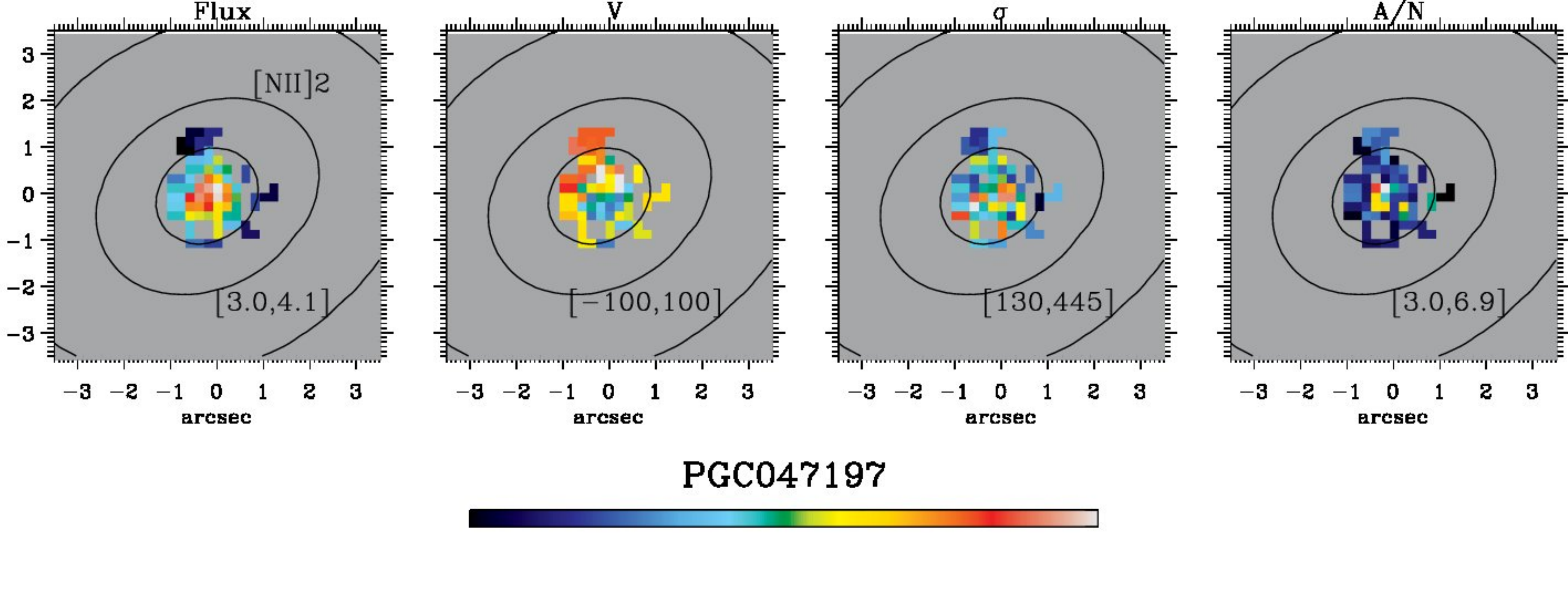} 
\caption{As in Fig.~\ref{fig:015524_1} but for PGC\,047197. 
Spatially resolved map of \niis 2.}
\label{fig:047197}
\end{figure*}


\begin{figure*}
\centering
\includegraphics[scale = 0.23, trim= 0cm 6cm 0cm 0cm, clip]{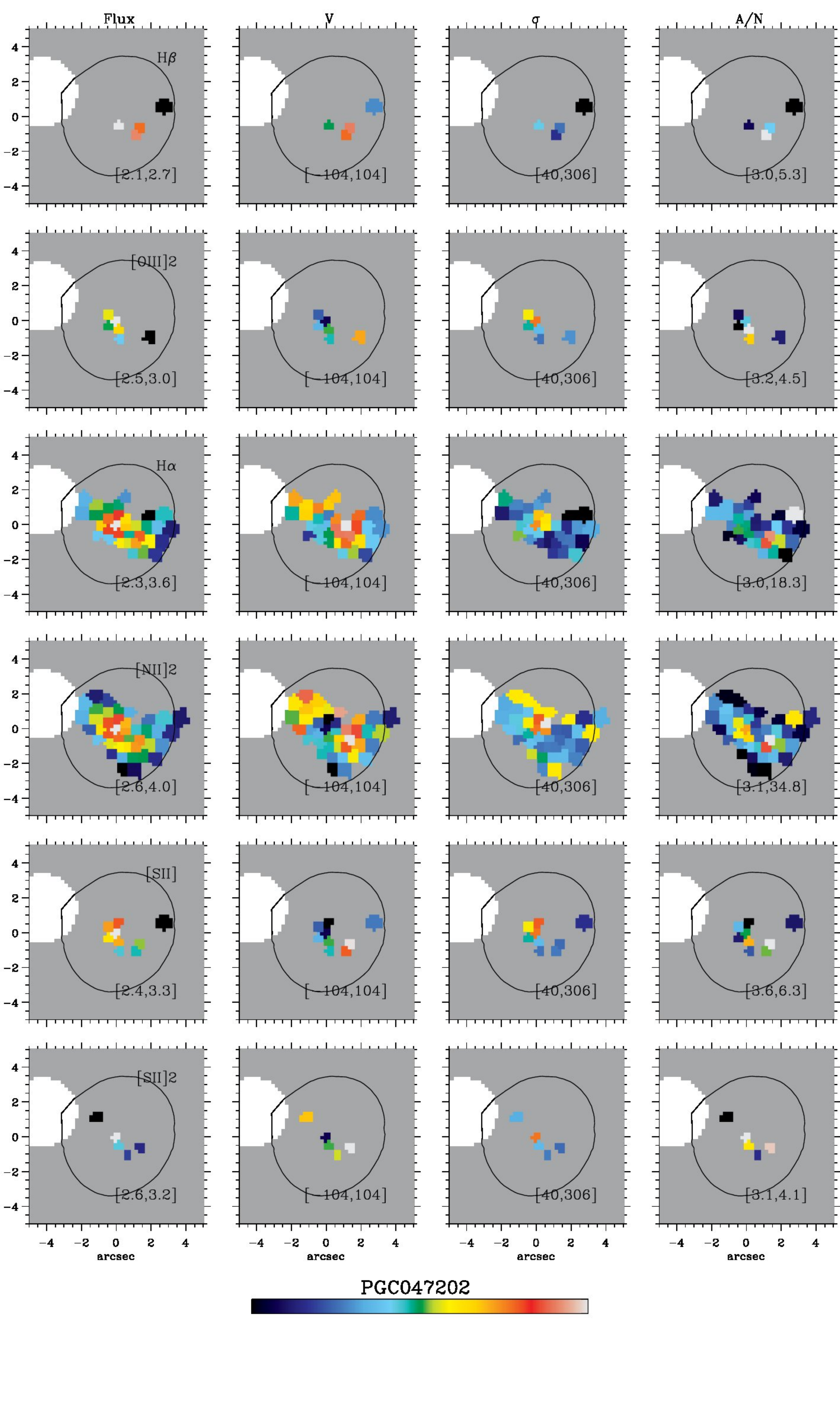} 
\caption{As in Fig.~\ref{fig:015524_1} but for PGC\,047202. 
{\em Up to down}: Spatially resolved maps of H$\beta$, \oiiis 2, H$\alpha$, 
\niis 2 and \siis\ and \siis 2.}
\label{fig:047202}
\end{figure*}


\begin{figure*}
\centering
\includegraphics[scale = 0.23, trim= 0cm 6cm 0cm 0cm, clip]{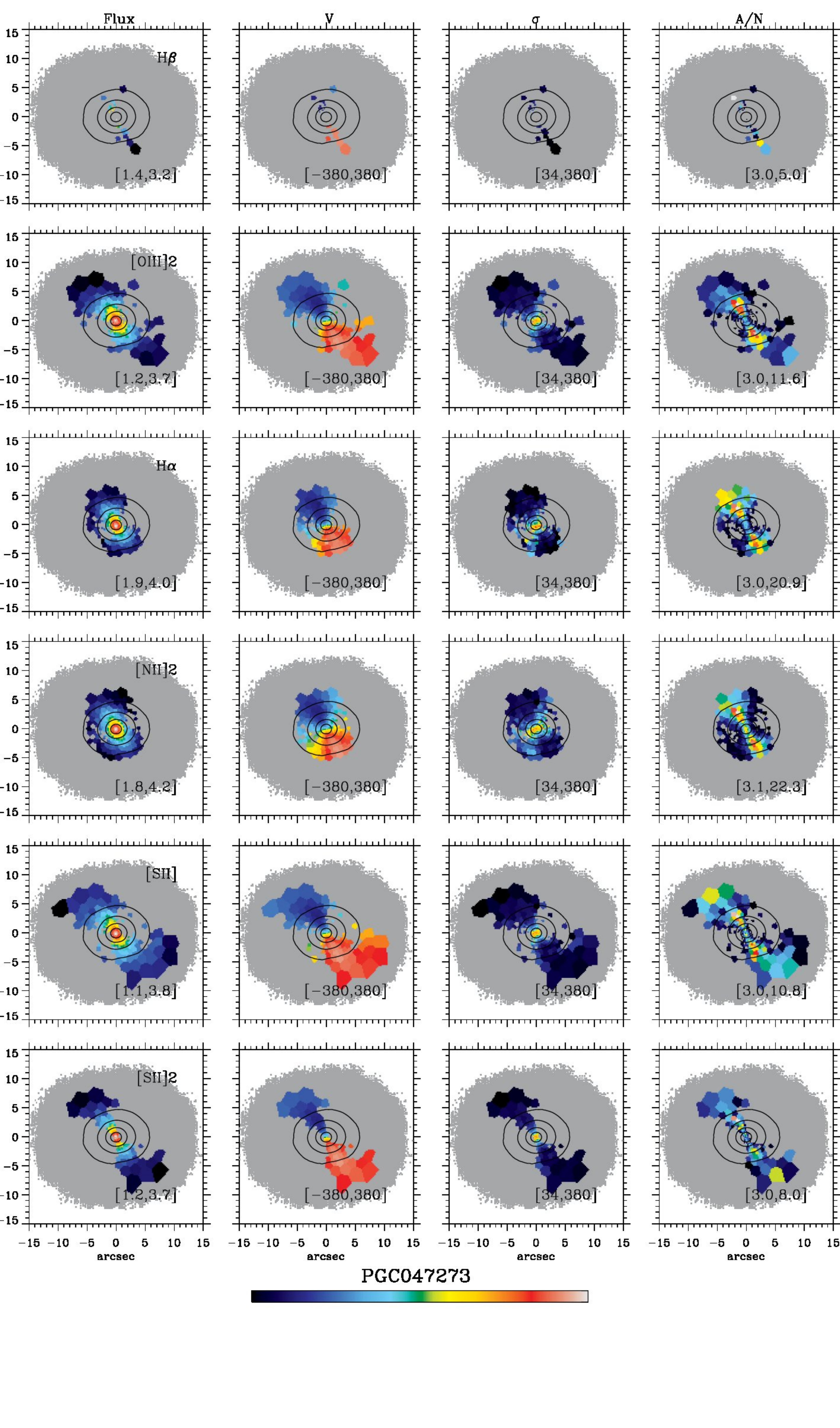} 
\caption{As in Fig.~\ref{fig:015524_1} but for PGC\,047273. 
{\em Up to down}: Spatially resolved maps of H$\beta$, \oiiis 2, H$\alpha$, 
\niis 2 and \siis\ and \siis 2.}
\label{fig:047273}
\end{figure*}


\begin{figure*}
\centering
\includegraphics[scale = 0.23, trim= 0cm 6cm 0cm 0cm, clip]{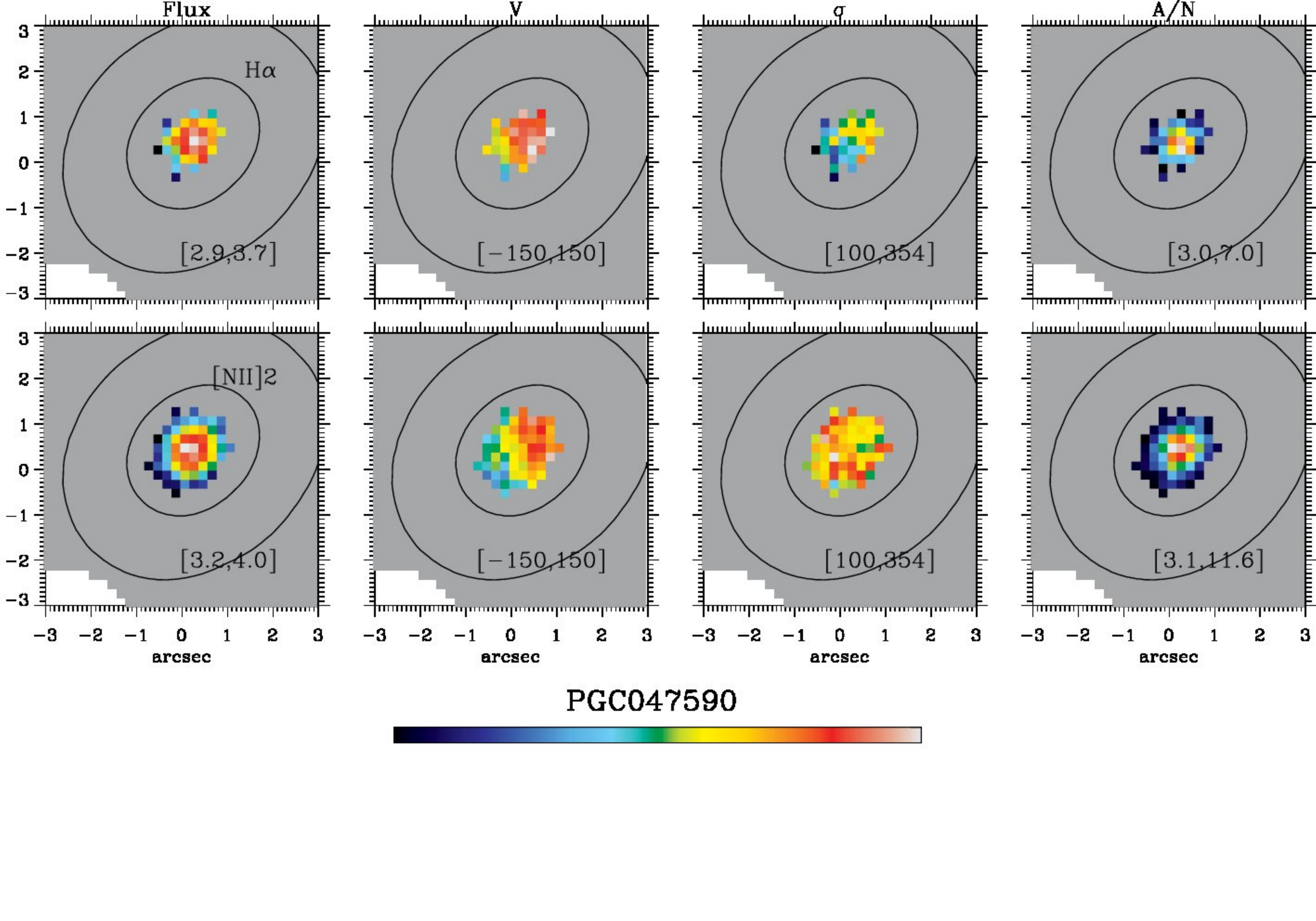} 
\caption{As in Fig.~\ref{fig:015524_1} but for PGC\,047590. 
{\em Up to down}: Spatially resolved maps of H$\alpha$ and \niis 2.}
\label{fig:047590}
\end{figure*}


\begin{figure*}
\centering
\includegraphics[scale = 0.23, trim= 0cm 7cm 0cm 0cm, clip]{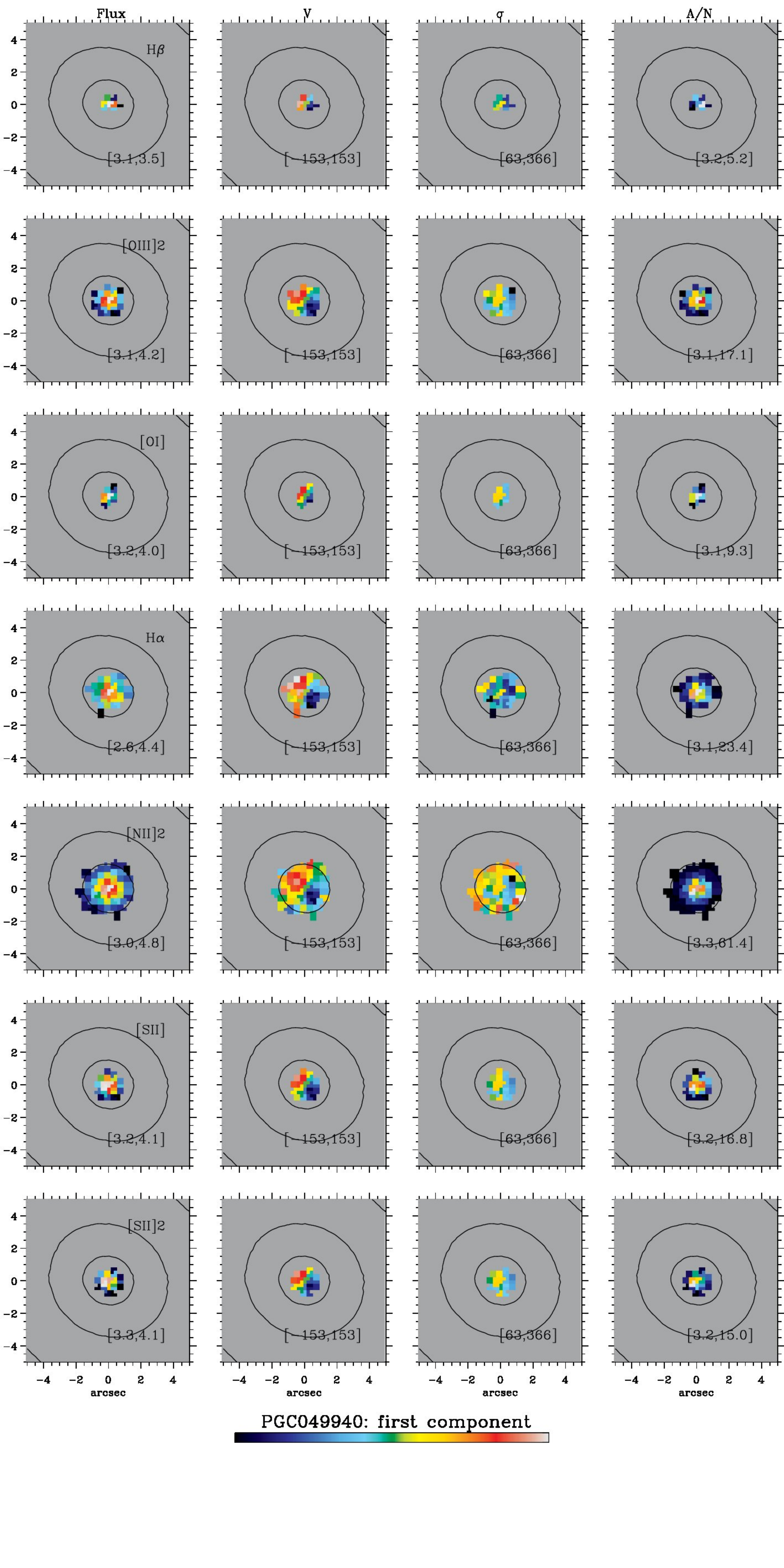} 
\caption{As in Fig.~\ref{fig:015524_1} but for the first velocity component of PGC\,049940. 
{\em Up to down}: Spatially resolved maps of H$\beta$, \oiiis 2, \ois, H$\alpha$, \niis 2 and \siis\ and \siis 2.}
\label{fig:049940_1}
\end{figure*}

\begin{figure*}
\centering
\includegraphics[scale = 0.23, trim= 0cm 7cm 0cm 0cm, clip]{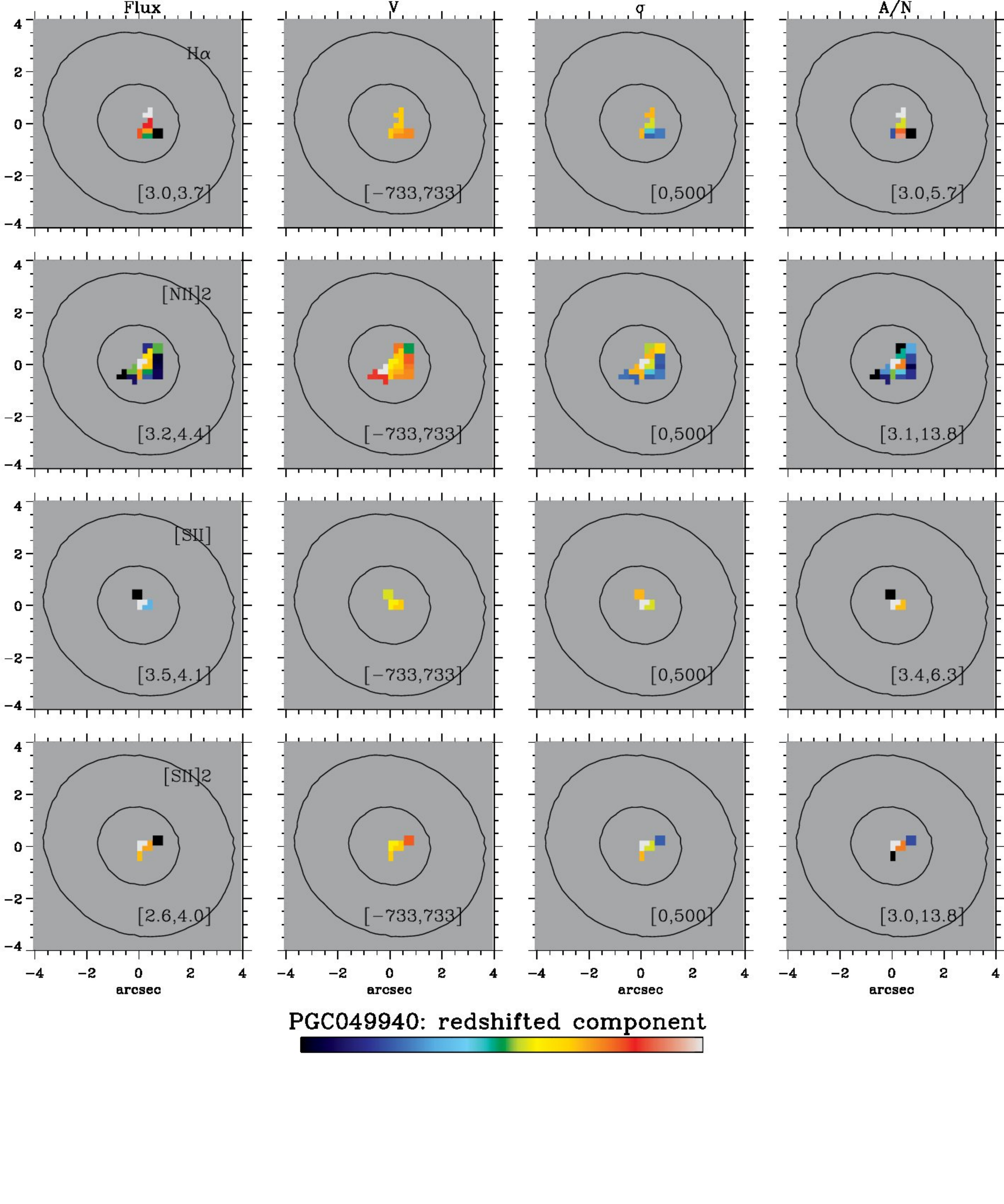} \\
\caption{As in Fig.~\ref{fig:015524_1} but for the redshifted component of PGC\,049940. 
{\em Up to down}: Spatially resolved maps of H$\alpha$, \niis 2 and \siis\ and \siis 2.}
\label{fig:049940_2}
\end{figure*}

\begin{figure*}
\centering
\includegraphics[scale = 0.23, trim= 0cm 25cm 0cm 14cm, clip]{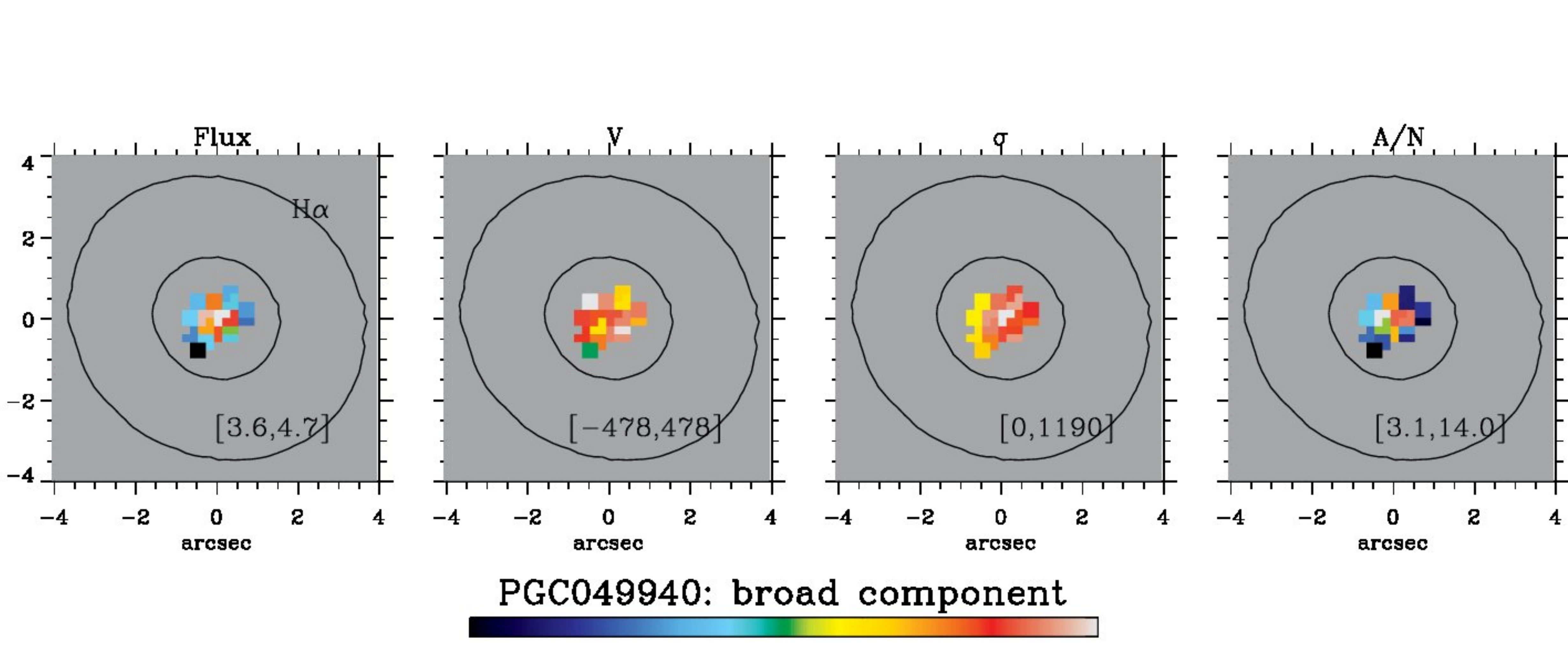} \\
\caption{As in Fig.~\ref{fig:015524_1} but for the broad H$\alpha$ component of PGC\,049940.}
\label{fig:049940_3}
\end{figure*}


\begin{figure*}
\centering
\includegraphics[scale = 0.23, trim= 0cm 6cm 0cm 0cm, clip]{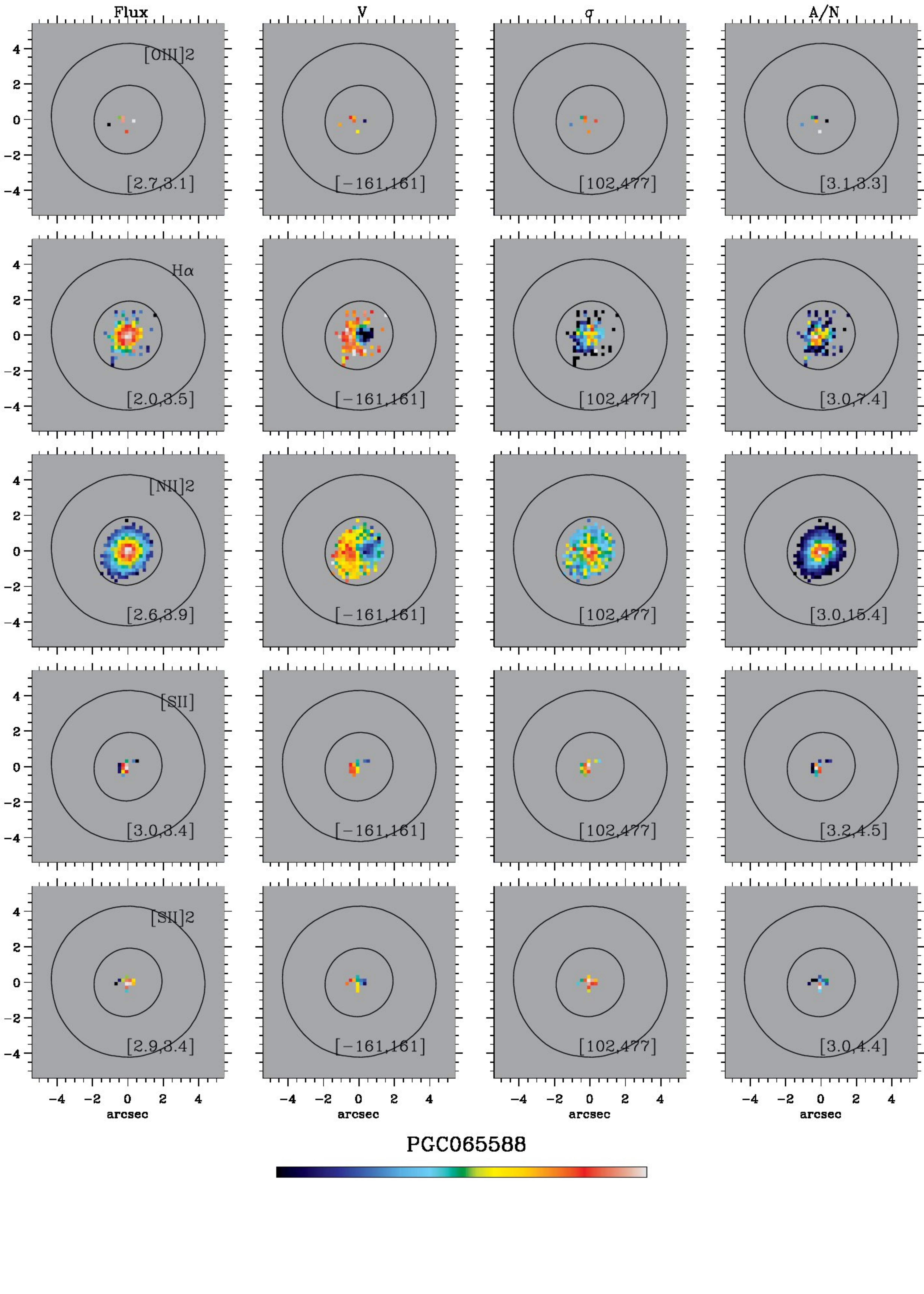} 
\caption{As in Fig.~\ref{fig:015524_1} but for PGC\,065588. 
{\em Up to down}: Spatially resolved maps of \oiiis 2, H$\alpha$, 
\niis 2 and \siis\ and \siis 2.}
\label{fig:065588}
\end{figure*}


\begin{figure*}
\centering
\includegraphics[scale = 0.225, trim= 0cm 6.8cm 0cm 0cm, clip]{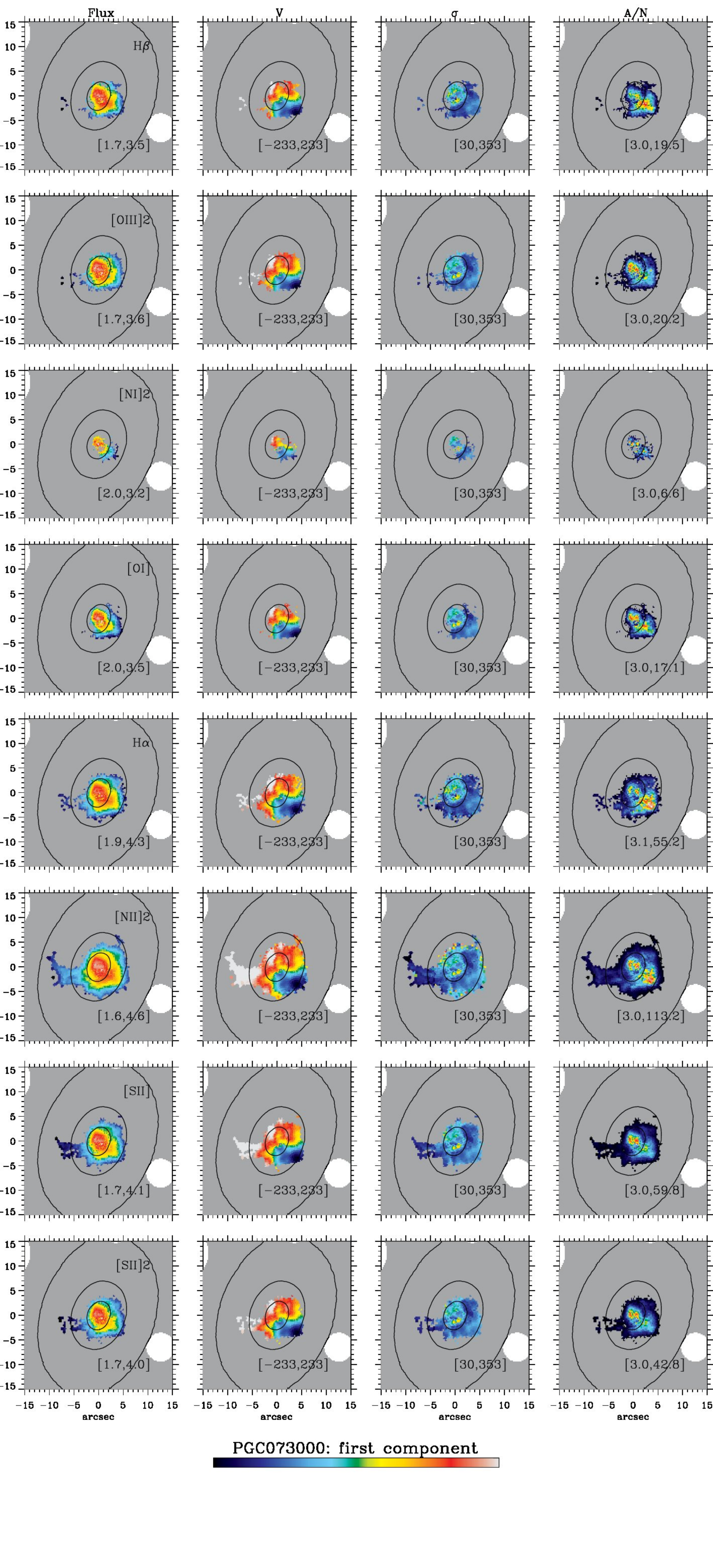} 
\caption{As in Fig.~\ref{fig:015524_1} but for the first velocity component of PGC\,073000. 
{\em Up to down}: Spatially resolved maps of H$\beta$, \oiiis 2, \nis 2, \ois, H$\alpha$, \niis 2 and \siis\ and \siis 2.}
\label{fig:073000_1}
\end{figure*}

\begin{figure*}
\centering
\includegraphics[scale = 0.225, trim= 0cm 6.8cm 0cm 0cm, clip]{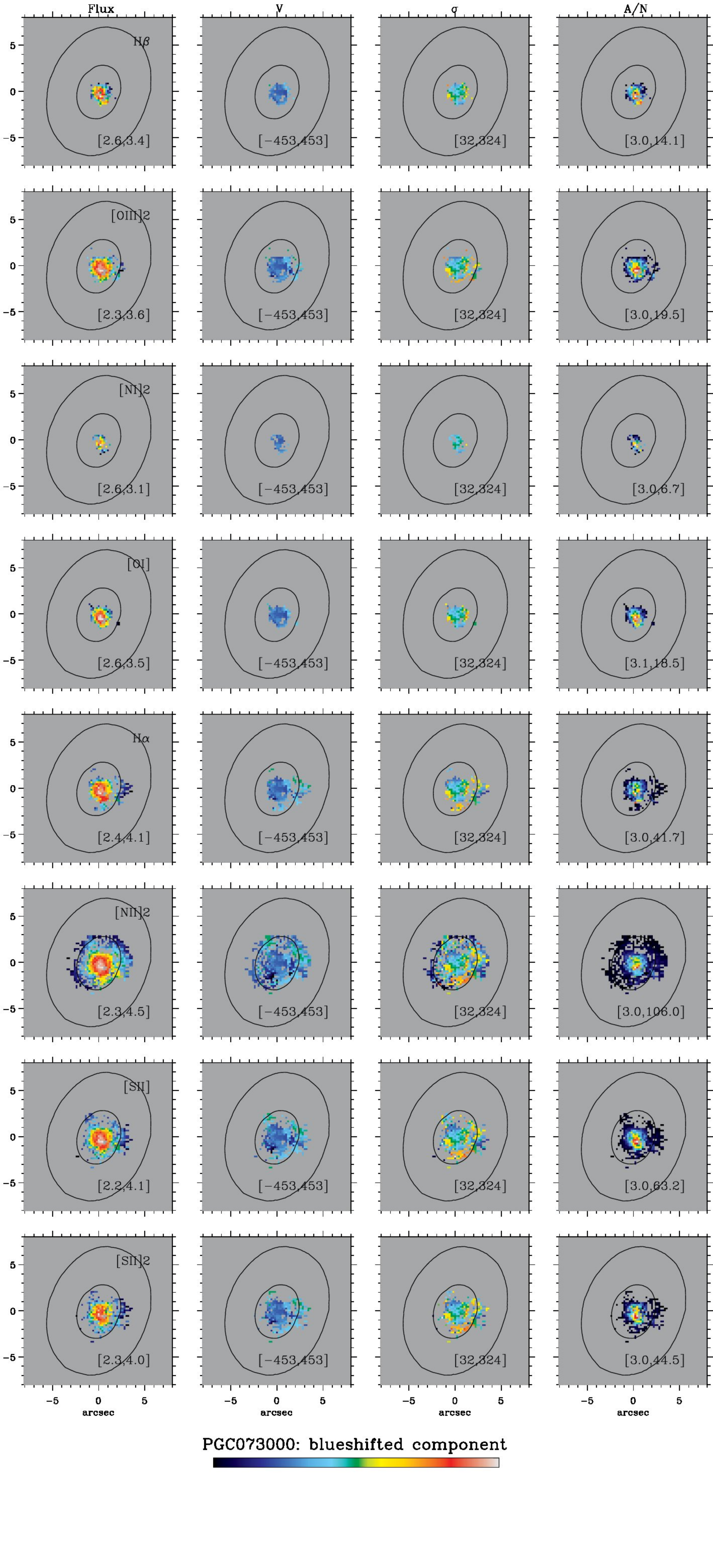} \\
\caption{As in Fig.~\ref{fig:015524_1} but for the blueshifted component of PGC\,073000. 
{\em Up to down}: Spatially resolved maps of H$\beta$, \oiiis 2, \nis 2, \ois, H$\alpha$, \niis 2 and \siis\ and \siis 2.}
\label{fig:073000_2}
\end{figure*}

\begin{figure*}
\centering
\includegraphics[scale = 0.23, trim= 0cm 18cm 0cm 0cm, clip]{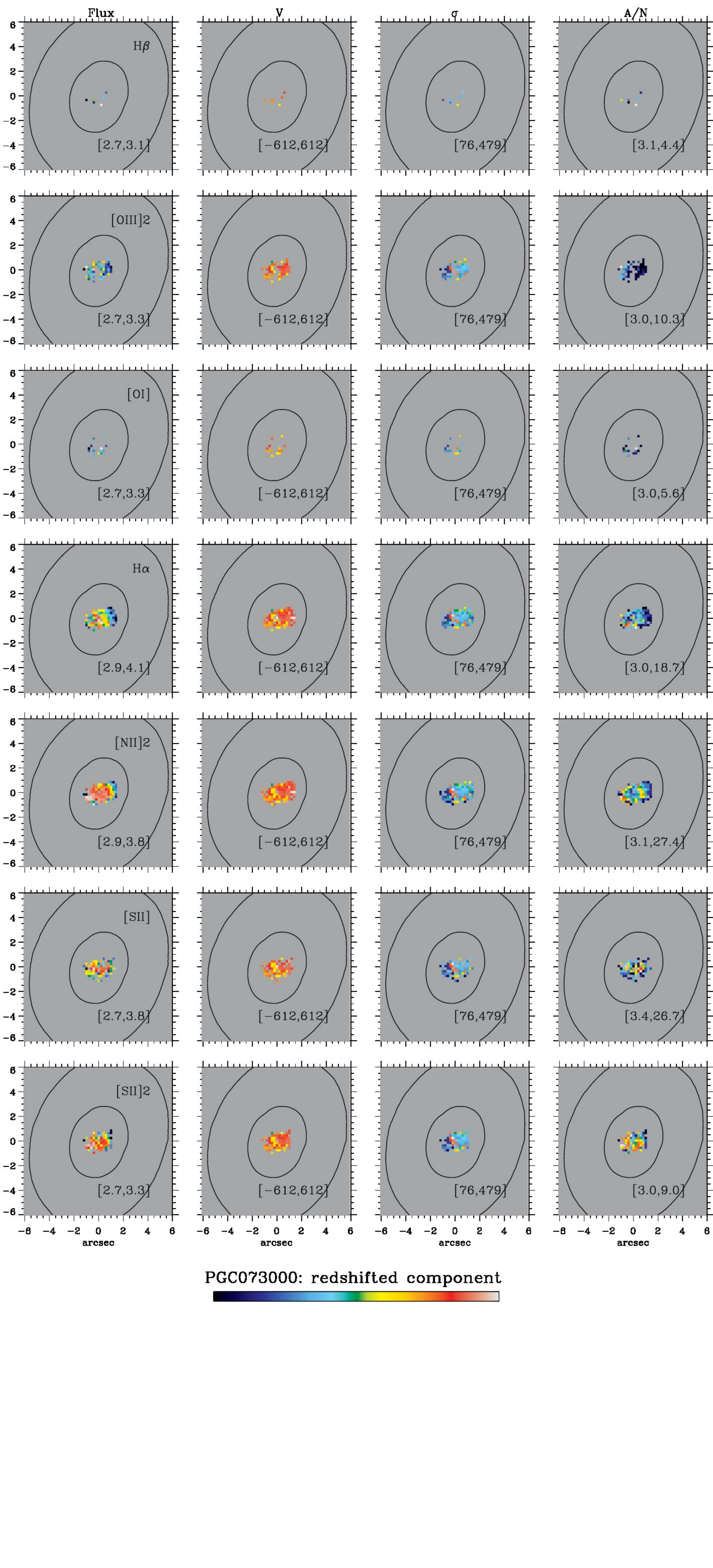} \\
\caption{As in Fig.~\ref{fig:015524_1} but for the redshifted component 
of PGC\,073000. {\em Up to down}: Spatially resolved maps of H$\beta$, \oiiis 2, \ois,
H$\alpha$, \niis 2 and \siis\ and \siis 2.}
\label{fig:073000_3}
\end{figure*}


\begin{figure*}
\centering
\includegraphics[scale = 0.23, trim= 0cm 18cm 0cm 0cm, clip]{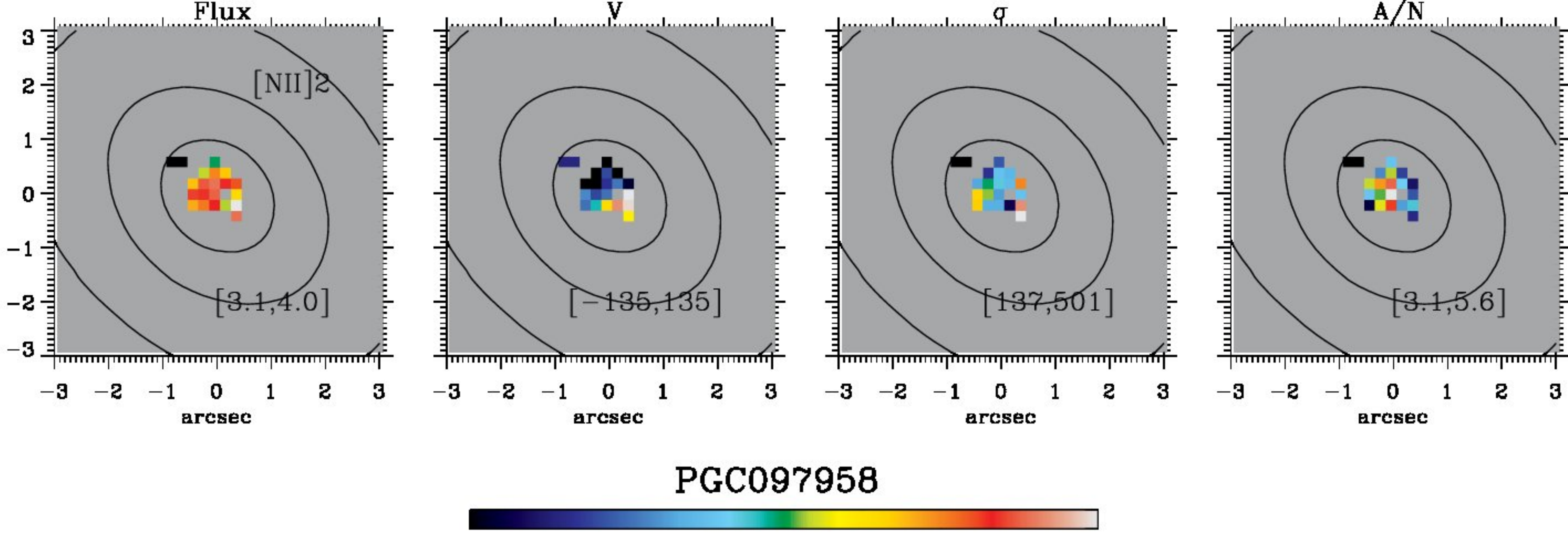} 
\caption{As in Fig.~\ref{fig:015524_1} but for PGC\,097958. 
Spatially resolved map of \niis 2.}
\label{fig:097958}
\end{figure*}

\end{appendix}
\end{document}